\documentstyle[12pt,aaspp4]{article}

\slugcomment{To appear in The Astrophysical Journal}

\begin{document}

\title{ Dense Gas in the Milky Way }

\author{Tamara T. Helfer\altaffilmark{1} and Leo Blitz\altaffilmark{1}}
\affil{Department of Astronomy, University of Maryland,
    College Park, MD 20742}

\altaffiltext{1}{thelfer,blitz@astro.berkeley.edu, current address Radio 
Astronomy Lab, 601 Campbell Hall, UC Berkeley, Berkeley CA 94720}

\begin{abstract}

We present a study of dense gas emission in the Milky Way in order
to serve as a basis for comparison with extragalactic results.  This
study combines new observations of HCN, CS, and CO in individual
GMCs and in the Milky Way plane with published studies of emission
from these molecules in the inner 500 pc of the Milky Way.  
We find a strong trend in the fraction of emission from dense gas 
tracers as a function of location in the
Milky Way:  in the bulge, $\rm I_{HCN}/I_{CO}$ = 0.081 $\pm$ 0.004,
in the plane, $\rm I_{HCN}/I_{CO}$ = 0.026 $\pm$ 0.008 on average,
and over the full extent of nearby GMCs, $\rm I_{HCN}/I_{CO}$ = 0.014 
$\pm$ 0.020.  Similar trends are seen in $\rm I_{CS}/I_{CO}$:
in the bulge, $\rm I_{CS}/I_{CO}$ = 0.027 $\pm$ 0.006,
in the plane, $\rm I_{CS}/I_{CO}$ = 0.018 $\pm$ 0.008 on
average, and over the full extent of nearby GMCs, 
$\rm I_{CS}/I_{CO}$ = 0.013 $\pm$ 0.02.   The
low intensities of the HCN and CS emission in the plane suggests
that these lines are produced by gas at moderate densities; they
are thus not like the emission produced by the dense, pc-scale
star forming cores in nearby GMCs.  The contrast between the
bulge and disk ratios in the Milky Way is likely to be caused by
a combination of higher kinetic temperatures as well as a higher
dense gas fraction in the bulge of the Milky Way.   We show that
the ratio $\rm I_{HCN}/I_{CO}$ is correlated quantitatively 
with the total hydrostatic gas pressure in the Milky Way as 
$\rm I_{HCN}/I_{CO} \propto P^{0.19\pm0.04}$.

\end{abstract}


\section{ Introduction }

Any study of extragalactic phenomena demands a comparison with
conditions in the Milky Way, which in the absence of
compelling contradictions is used as a reference for what
is ``normal'' in a spiral galaxy. 
The result that the centers of most nearby spiral galaxies
contain a large amount of gas at densities of $\sim 10^5$ cm$^{-3}$
(traced by HCN and CS emission, Nguyen-Q-Rieu et al. 1992; 
Helfer \& Blitz 1993) is surprising when 
compared to conditions in giant molecular clouds (GMCs) in the solar 
neighborhood, where typical clump densities are only $\sim 
10^{2.5} - 10^{3.5}$ cm$^{-3}$ (Blitz 1987) and which contain relatively 
few clumps of dense gas emission (e.g. Lada, Bally, \& Stark 1991).  
What is questionable, however, is how meaningful such a comparison
between the centers of galaxies and the local environment of the
Sun is to the interpretation of the HCN and CS observations.  The
goal of this paper is to investigate the properties of 
GMCs as a function of their location 
within the Milky Way: do GMCs within the high-pressure environment 
of the Galactic bulge have the same physical properties as GMCs 
in the Galactic disk and near the Sun?   Throughout this study, we
use as a qualitative indicator of pressure the ratios of the 3 mm 
emission from the dense gas tracers HCN and CS relative to that of CO.  
(HCN and CS typically trace gas densities of 10$^5$ cm$^{-3}$ or higher; 
this is about 100 times the critical density required to excite CO.)

In the Milky Way, emission from dense molecular gas tracers
has been studied in the bulge (CS: Bally et al. 1987; HCN: Jackson
et al. 1996), where it is distributed over a size scale of
$d \sim$ 500 pc, and in many individual star-forming cores in nearby
GMCs, on size scales of a few pc or smaller.  To complicate 
extragalactic comparisons with the Milky Way, single-dish 
radio beams typically cover size scales of tens
to hundreds of GMC-sized diameters in nearby galaxies, and
even interferometric beams correspond to size scales of a
few GMCs.  (A GMC has a typical diameter of 50 pc; Blitz 1987.) 
The disparate size scales over which these observations have been
made preclude a fair comparison of these different regions.  In this
paper, we place ourselves at the telescope of an imaginary observer in
a nearby galaxy, and we present a systematic study of molecular cloud 
properties in the Milky Way on size scales of GMCs and larger.

For the bulge of the Milky Way, we use the studies of Bally et al. (1987)
(as presented in Helfer \& Blitz 1993) and Jackson et al. (1996).  We also 
present new observations of
the disk of the Milky Way, which are divided into two types.  First,
we observed five local GMCs, one high-latitude cloud, and a distant
outer-galaxy molecular cloud over their entire extents in
CO, HCN and CS, in order to make the first measurements of dense gas 
ratios over a sample of whole clouds.  Second, we surveyed some 40\arcdeg\ in
the first quadrant of the Milky Way using an unbiased sampling in order 
to determine the dense gas 
ratios in the Galactic plane, where the number density of molecular
clouds is relatively high compared with the local environment.  

Throughout this study, we assume that the Sun lies at a distance of
R$_{\rm o}$ = 8.5 kpc from the Galactic Center.

\section{ Line Ratios as Tracers of High-Pressure Gas }

In this study, we use as a qualitative measure of pressure the 
line ratios of the 3 mm transitions of HCN and CS to that of
CO.  Here, we examine the excitation characteristics
of these molecules in order to justify the use of these ratios as
indicators of gas pressure.  Since much of this study as well
as current extragalactic research emphasizes
HCN observations over CS observations, we consider the excitation
of the HCN molecule in the following discussion.

For CO, the J = 1 state lies only 5.5 K above the ground
rotational state, and for HCN, the difference is only 4.3 K;
this means that even rather cold gas has the energy
to populate the J = 1 state for both molecules.  Since the
population of the upper rotational states is accomplished
primarily by collisional excitations, then to first
order, the H$_2$ density plays a more important role than 
does the kinetic temperature in the excitation of these molecules.
Furthermore, since the CO molecule has a relatively low dipole
moment compared to HCN (0.1 debye for CO, 3.0 debye for
HCN), it takes much higher densities to excite the HCN molecule:
the critical excitation density of HCN is about 10$^3$ times
higher than that of CO.

\begin{figure}
\plotfiddle{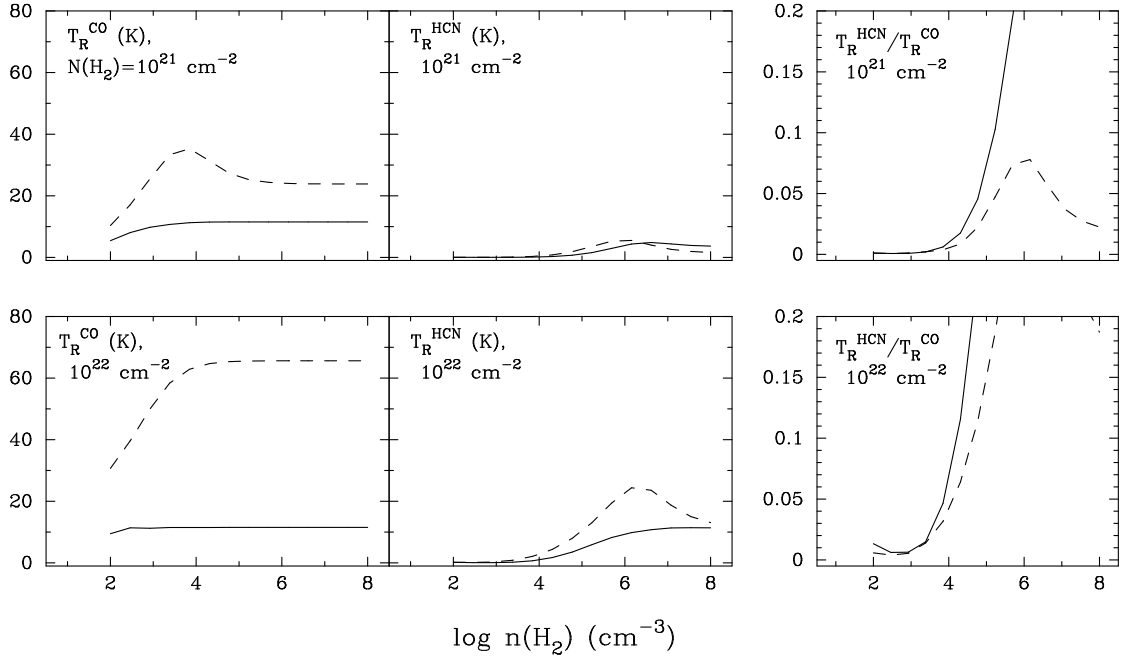}{300pt}{-90}{57}{57}{-230}{360}
\caption{Results from sample LVG calculations of CO and HCN line 
excitation.  Shown are results for the J = 1 -- 0 transitions only.  
The solid lines are for T$_{\rm K}$ = 15 K, and the dashed lines are 
for T$_{\rm K}$ = 70 K.  
The calculations assume [HCN]/[CO] = 10$^{-4}$ and [CO]/[H$_2$]
= 8 $\times$ 10$^{-5}$. The top row is for N(H$_2$) = 
1.3 $\times$ 10$^{21}$ cm$^{-2}$, and the bottom is for 
N(H$_2$) =  1.3 $\times$ 10$^{22}$ cm$^{-2}$.   The left panels show the 
CO radiation temperature in K, the middle panels show the HCN
radiation temperature in K, and the right panels show the ratio
of the radiation temperatures T$_{\rm R}$(HCN)/T$_{\rm R}$(CO).
}
\label{lvgfig}\end{figure}

With this general picture in mind, we ran model calculations
using a large velocity gradient (LVG) radiative transfer code 
(provided by L.G. Mundy)
to examine the excitation properties of the J = 1 -- 0 transitions 
of the HCN and CO molecules.  We assumed the relative
abundances [HCN]/[CO] = 10$^{-4}$ and [CO]/[H$_2$] = 8 $\times$
10$^{-5}$ for these calculations and ran models for gas at 
kinetic temperature of 15 K and 70 K. The models assume a velocity
gradient of 1 km s$^{-1}$ pc$^{-1}$. The results of the calculations
are summarized in Figure \ref{lvgfig}. The top row shows the
results for a total H$_2$ column density N(H$_2$) = 1.3 $\times$
10$^{21}$ cm$^{-2}$, and the bottom row is for N(H$_2$) =
1.3 $\times$ 10$^{22}$ cm$^{-2}$.  As the left panels show,
the CO J = 1 state is already reasonably well populated at densities
of n(H$_2$) = 10$^2$ cm$^{-3}$, and there is little variation
in the line strengths as a function of density.  By contrast, the middle 
panels show that, for densities below $\la$ 10$^{4-5}$ cm$^{-3}$,
the radiation temperatures for the HCN J = 1 -- 0 transition are small;
the biggest increase in line temperature occurs over the density interval 
$\sim$ 10$^{3.5}$ cm$^{-3}$ -- 10$^{6}$ cm$^{-3}$.  

The line ratios T$_{\rm R}$(HCN)/T$_{\rm R}$(CO) are shown in the
right panels of Figure \ref{lvgfig}.  These panels show the general 
trend that for low temperature gas as well as for high temperature
gas at densities $\la$ 10$^6$ cm$^{-3}$, a higher line ratio
T$_{\rm R}$(HCN)/T$_{\rm R}$(CO) implies a higher density gas.
Thus for gas at the center of the Milky Way, where T$_{\rm K}
\sim$ 70 K and N(H$_2$) $\sim$ 10$^{22}$ cm$^{-2}$, the measured
HCN to CO line ratios of 0.05 -- 0.2 (Jackson et al. 1996; see $\S$ 5) 
imply densities of 10$^4$ -- 10$^{5.3}$ cm$^{-3}$, whereas the
ratios of $\sim$ 0.02 -- 0.03 measured in the Milky Way disk
($\S$ 4.3), where T$_{\rm K} \sim$ 15 K and N(H$_2$) $\sim$ 
10$^{22}$ cm$^{-2}$, imply mean densities of $\sim 10^{3.5}$ cm$^{-3}$.
In summary, although the exact results implied by the LVG analysis 
are dependent on an accurate knowledge of a number of variables, 
including the column densities, abundances, kinetic temperatures 
and velocity gradients -- for which there is no independent measure
in this study --  the general 
consistency between the known conditions in the Milky Way and the ratios 
measured in the Milky Way bulge and disk lend credibility to the use of
these line ratios as a qualitative measure of the pressure in
galaxies.

\section{ Solar Neighborhood GMCs}

The clouds surveyed are listed in Table 1.  The five sources that
form the core of this study are the S88, S140, S269, Orion B, and Rosette
molecular clouds; these are GMCs that lie within 4 kpc of the 
Sun.  The source MBM 32 is a high-latitude molecular cloud, rather than 
a GMC; its mass is much lower than that of a typical GMC ($\rm M_{MBM32} 
\sim 100 ~M_{\sun}$, while GMCs are typically taken to be clouds with
mass $\rm \ga 10^4$ or $\rm 10^5 ~M_{\sun}$, Blitz 1993), 
and it is not self-gravitating (Magnani, Blitz, \& Mundy 1985).  The 
source OGC2 is an distant outer-galaxy molecular cloud; at a heliocentric 
kinematic distance of 21 kpc, this cloud lies beyond the optical disk of 
the Milky Way.

\subsection{ FCRAO 14 m Observations }
We observed the clouds S88, S140\footnote{CO map kindly provided by
M.H. Heyer}, S269, Orion B, and MBM 32 using the
Five Colleges Radio Astronomy Observatory (FCRAO) 14 m telescope in New 
Salem, MA in 1993 March 17-23, May 17-23, and October 22-27.    We used 
the QUARRY multibeam receiver, which is an array receiver of 3 $\times$ 5
Schottky diode mixers; this allowed for fast mapping of large fields with good 
sampling in a reasonable amount of time.  At that time, the backends 
were 15 32-channel 
spectrometers with a resolution of 1 MHz; at 100 GHz, this corresponds to a 
velocity resolution of 3.0 km s$^{-1}$ over a bandwidth of 96 km s$^{-1}$.
Typical system temperatures were 700 -- 1500 K at 115 GHz and 400 -- 900 K
at 88 GHz and 98 GHz.  The FWHM of the individual QUARRY beams is about 
55\arcsec\ at 100 GHz; at a canonical cloud distance of 2 kpc, this corresponds
to a linear resolution of about 0.5 pc.

The goal was to simulate an observation of each source with a telescope that 
has a beam large enough to cover the entire extent of the cloud as defined
by the CO emission, typically $r$ = 20 pc.
The general observing strategy was to cover the entire extent of the
cloud with 30 to 60 second integrations at each position in each of the
species CO J = 1-0 (115.27 GHz), CS J = 2-1 (97.98 GHz), and 
HCN J = 1-0 (88.63 GHz).  
For most of the clouds observed, we sampled at full beamwidth spacing. In the 
case of Orion B, we sampled about one quarter of the cloud area with sparser
sampling.  The resulting maps include some 2300 spectra in the case of
S269, all the way up to 23000 spectra for S140.  

Since the large data sets precluded a visual inspection of all the 
individual spectra, we reduced the spectra using automated routines.  The data
were transferred from the FCRAO SPA format into the AT\&T Bell Laboratories
package COMB for reduction.  First, the spectra were scanned automatically
for channels with absolute values that were outside a given tolerance.
For a given spectrum, if there were two adjacent channels which were
outside the tolerance, or if there were more than three isolated channels
that were outside the tolerance, then the spectrum was rejected from
the data set.  Otherwise, any isolated bad channels were overwritten with
the average of the two adjacent channels (or, if the bad channel was at
the edge of the spectrum, the bad channel was replaced by the value of
its neighboring channel).  A linear baseline\footnote{In the case of the 
Orion B CS and the S269 HCN and CS data, there were low-amplitude standing 
waves apparent in the averaged spectra which required a fourth-order 
polynomial fit to the baseline to correct.}  was then fit to each spectrum
from channels outside the velocity limits of the line emission.  We made 
maps of integrated intensity using a ``cone interpolation'' smoothing scheme 
that weights each pixel by ($1 - {r \over r_i}$), where $r_i$ is the 
interpolation radius, typically chosen to be 1\arcmin\ (but chosen to be 
9\arcmin\ for the case of Orion B).  A visual inspection of the maps allowed 
us to go back and eliminate corrupt spectra that had been missed in the 
automated reduction. 

The data cubes were then transferred to the MIRIAD package (Sault, Teuben,
\& Wright 1995) for further processing.  We used channel maps of the CO data
to determine the velocity extent of emission in the clouds, and we made
maps of integrated intensity in CO by summing over those channels with 
detected emission.
Maps of CS integrated intensity were made with the same velocity coverage
as that of the CO.  The HCN molecule has triplet fine-structure lines
in its J = 1-0 transition; the limits of integration for the HCN integrated
intensity maps were therefore extended by -7 km s$^{-1}$ and +5 km s$^{-1}$ 
to account for the fine structure lines\footnote{In the case of
the S269 HCN data, we did not use the full extent of the fine-structure
velocity range in the limits of integration.}.  Finally, we created a mask
of the CO, HCN, and CS maps to ensure that only pixels that had been
sampled in all of the three species were included in the subsequent
analysis.

\subsubsection{ The FCRAO Maps }

\begin{figure}
\figurenum{2 continued}
\plotfiddle{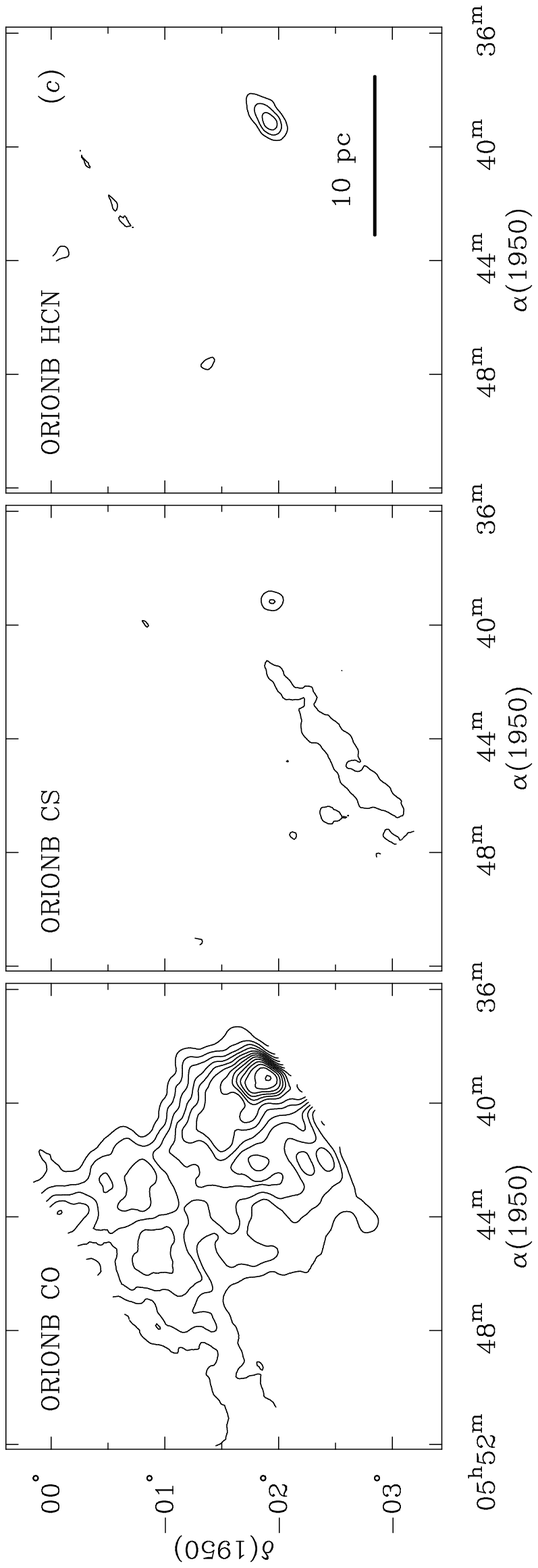}{360pt}{-90}{70}{70}{-255}{580}
\vspace*{-1in}
\plotfiddle{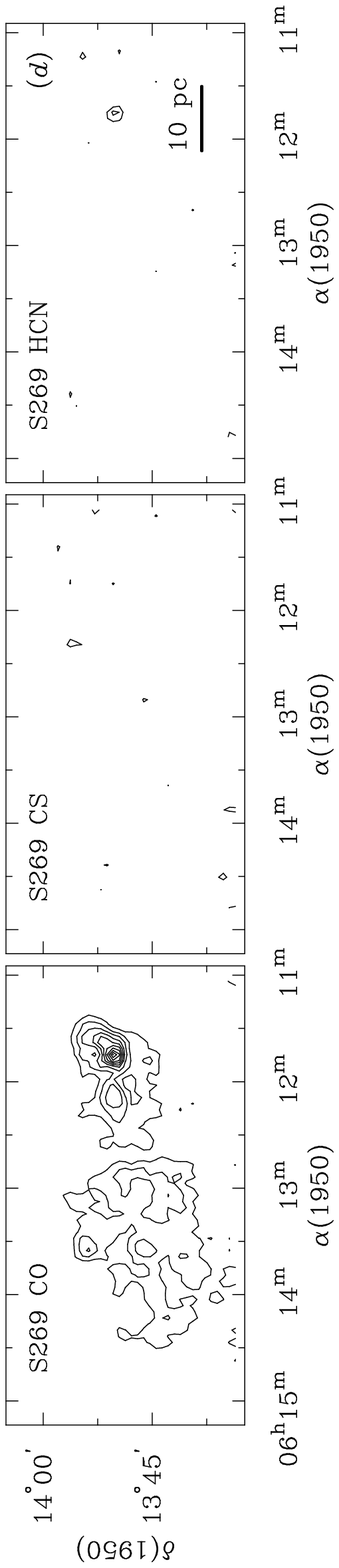}{-180pt}{-90}{70}{70}{-255}{400}
\plotfiddle{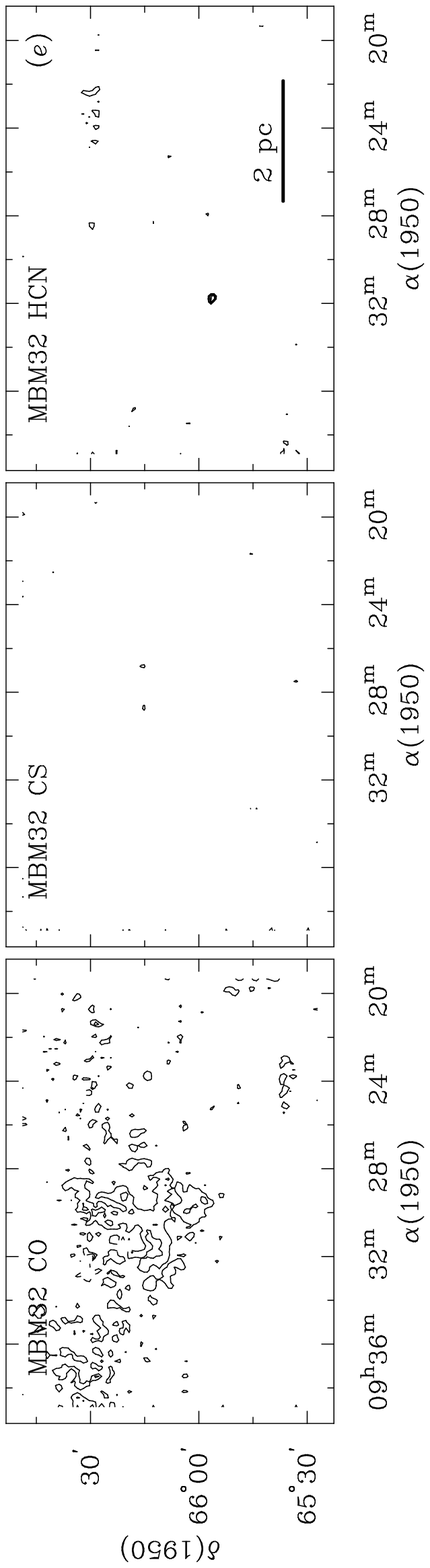}{-180pt}{-90}{70}{70}{-255}{300}
\caption{ ($c$) Same for Orion B.
The CO contours are as above; the CS contours are $\pm$ 1.5,3,4.5...
K km s$^{-1}$; and the HCN contours are $\pm$ 2.1,4.2,6.3... K km s$^{-1}$.
($d$) Same for S269.  The CO
contours are as above; the CS contours are $\pm$ 2.4,4.8,7.2... K km s$^{-1}$;
and the HCN contours are $\pm$ 2.1,4.2,6.3... K km s$^{-1}$.
($e$) Same for MBM32.  The
CO contours are $\pm$ 3,6,9... K km s$^{-1}$; the CS contours
are $\pm$ 1.2,2.4,3.6 K km s$^{-1}$; and the HCN contours are
$\pm$ 2.1,2.8,3.5... K km s$^{-1}$.}
\label{fcraomaps}\end{figure}

The FCRAO maps are presented in Figure \ref{fcraomaps}.  The
CO emission from S140, for example, is distributed over some 40 pc 
which defines the extent
of the GMC; however, the peak of the emission is strongly concentrated 
at the well-studied starforming core ($\alpha (1950) = 22^h 17^m 42^s, 
\delta (1950) = 63\arcdeg 04\arcmin$).   Figure \ref{fcraomaps} shows 
the dramatic contrast between the extended CO emission and the emission from 
HCN and CS, which at the sensitivity of these observations is 
confined to the central $\sim$ 2 pc around the S140 core; all the clouds show
the paucity of dense condensations in GMCs and demonstrate how little
of the volume of a GMC contains cores with the physical conditions required
to produce HCN and CS lines of substantial strengths.
What we wish to determine from these observations is whether there is 
low-level emission, perhaps not detected in a single spectrum, but detected 
in the average of hundreds or thousands of spectra whose positions are 
located throughout the GMC, that contributes positively to the total 
integrated emission of the cloud.

It is worthwhile pointing out once again the challenge of observing
objects of different size scales with modern single-dish telescopes
and interferometers.  The 1\arcmin\ beam size of the observations
is shown in the upper right corner of the S140 CO image in 
Figure \ref{fcraomaps}.  Although this beam size is tiny compared 
to the size of the region mapped, it is comparable to the 
{\it primary} beam size of modern interferometers!   At the distance of
a nearby galaxy, e.g. $d$ = 5 Mpc, the same beam size corresponds to
a size scale of 1500 pc.  Clearly, the task of making meaningful comparisons
of these disparate objects is exacerbated by the gross differences in the
linear resolutions of their observations.

\subsubsection{ Further Processing of the FCRAO Data }

Since the GMCs we mapped are not circularly symmetric, we
developed an automated routine to measure the total CO, HCN, and CS
fluxes and their ratios as a function of a threshold of CO
emission, rather than as a function of the physical distance
from the cloud core.  The threshold of CO
emission was characterized by multiples of the uncertainty in the
CO integrated intensity map (see below) as $\rm I_{clip}$ = 0,3,6,9...
$\times$ $\sigma_{\rm CO}$.  For each clipping level, a CO map was
generated from the original integrated intensity map by including only 
those pixels whose absolute value exceeded $\rm I_{clip}$.  This new CO map 
was then used as a template with which to select the corresponding pixels 
in the HCN and CS maps.

\begin{figure}
\plotfiddle{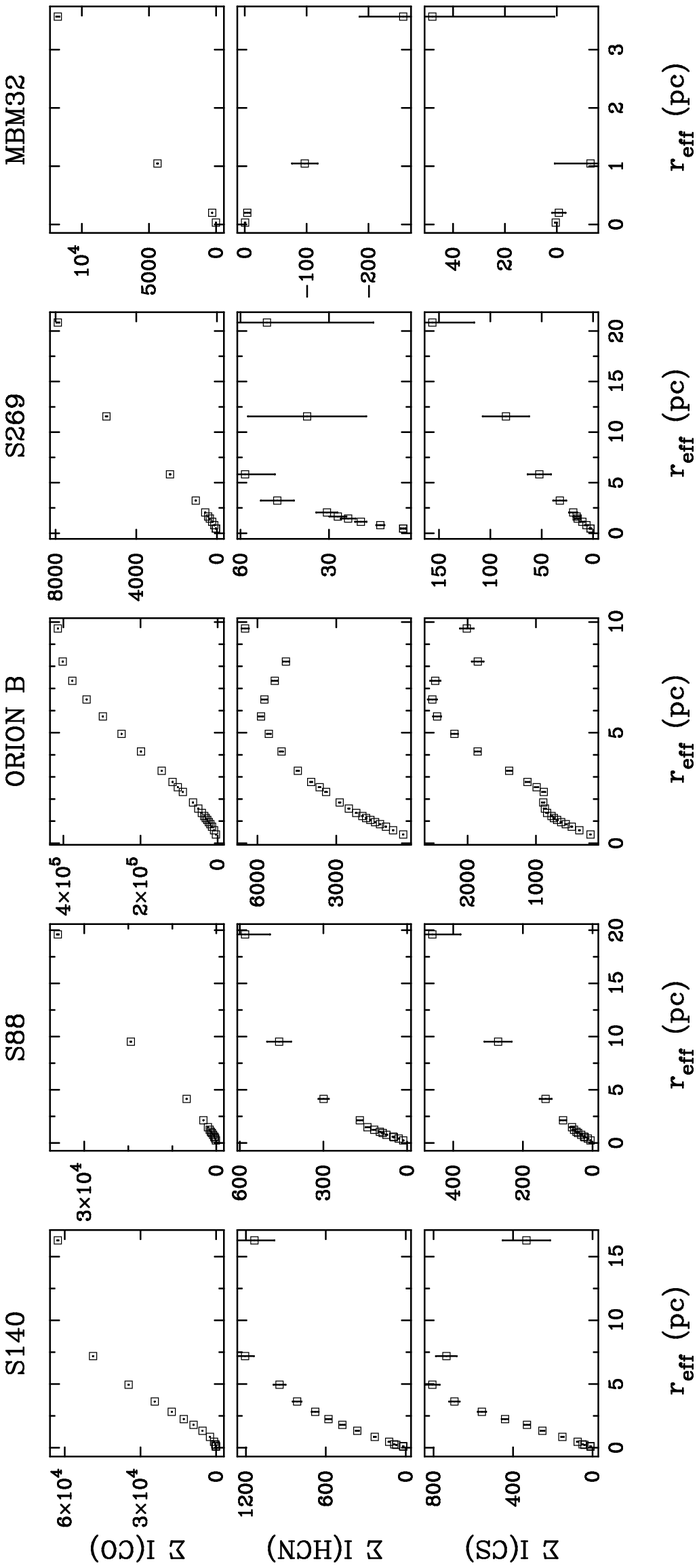}{460pt}{-90}{70}{80}{-260}{500}
\caption{The summed intensities $\Sigma$ I(CO), $\Sigma$ I(HCN), and 
$\Sigma$ I(CS) as a function of effective radius in each of the five 
clouds observed with QUARRY. The units of the ordinates are
K~km~s$^{-1}$. The error bars underestimate
the systematic uncertainties in the ratios at r$_{\rm eff}$ $\ga$ 
5 pc (see text).}
\label{fcraotot}\end{figure}

The results of this processing for S140, S88, Orion B, S269, and MBM32
are presented in Figure \ref{fcraotot}.  The threshold levels of CO, 
$\rm I_{clip}$, are represented as an effective radius $\rm r_{eff}$ 
in each cloud, where $\rm r_{eff} = 0.5 \sqrt{N} d$ $\Delta x$; N is 
the number of pixels included in each map (where the clipping level was 
set by $\rm I_{clip}$), d is the distance to the cloud, and $\Delta x$ 
is the angular extent of an individual pixel.   The analysis may thus
be thought of as a kind of ``aperture photometry'' so that
Figure \ref{fcraotot} represents the summed intensities of CO, HCN, and
CS as a function of the effective radius of the aperture.  
We note that near the core of each cloud, the effective radius is the same
as the physical radius; the physical and effective radii diverge at
distances of a few parsecs.

In Figure \ref{fcraotot}, the summed CO intensity $\Sigma$ I(CO)
rises monotonically as a function of the effective radius for each 
the clouds observed. This is easily understood, since the cloud is
defined by its CO extent and more and more of the cloud is included 
in the sum as the effective radius increases.  For the HCN and
CS emission, we expect that the summed intensities should also rise
monotonically, but that at some $\rm r_{eff}$ they might reach 
some plateau that represents the total emission in HCN or CS.
As Figure \ref{fcraotot} shows, this is the case for e.g. the
S140 HCN panel and the S88 panels.  But Figure \ref{fcraotot} also
shows that the total intensity {\it decreases} for some of
the clouds at $\rm r_{eff} \ga$ 5 pc, e.g. for S140 CS and for Orion 
B HCN and CS.  These decreases
are not physical; they are representative of systematic errors
that appear at the highest sensitivities in the maps (see below).
Figure \ref{fcraotot} also shows that low-level emission from
HCN and CS generally contributes positively to the summed
intensities at least out to effective radii of $\sim$ 5 pc.

The ratios $\rm I_{HCN}/I_{CO}$ and $\rm I_{CS}/I_{CO}$ are shown as a 
function of the effective radius of the aperture in Figure \ref{panel}.
As this figure shows, the ratios are strong functions of the
effective radius $\rm r_{eff}$.  The peak values for $\rm I_{HCN}/I_{CO}$
range from 0.1 -- 0.3 for all four clouds on size scales of $\rm r_{eff}
\le 1$ pc, but by $\rm r_{eff}$ = 5 pc, $\rm I_{HCN}/I_{CO}$ has
dropped to 0.04 or below, and by $\rm r_{eff} \ge$ 10 pc,  
$\rm I_{HCN}/I_{CO}$ is $\le$ 0.02 for all four clouds.  The ratio
$\rm I_{CS}/I_{CO}$ shows a similar trend.

The average ratios over the entire clouds are $<$$\rm 
I_{HCN}$$>$/$<$$\rm I_{CO}$$>$ = 0.007 -- 0.019 and $<$$\rm 
I_{CS}$$>$/$<$$\rm I_{CO}$$>$ = 0.006 -- 0.020.
For MBM 32, neither HCN nor CS was detected; the 3$\sigma$ upper
limits were $<$ 0.015 for $<$$\rm 
I_{HCN}$$>$/$<$$\rm I_{CO}>$ and $<$ 0.009 for $<$$\rm 
I_{CS}$$>$/$<$$\rm I_{CO}$$>$.

\subsubsection{ Limitations of the Observations }

The experiment that we have described above, i.e. that of making high 
sensitivity measurements of low-level emission, is a particularly
challenging one to do with good accuracy at millimeter wavelengths.  
At high sensitivities, systematic uncertainties dominate the random errors and
one must pay special attention to problems like low-amplitude
standing waves in the telescope spectrometer (see footnote 4
in $\S$ 3.1) and the details of the baseline fitting.  In these
QUARRY observations, the spectrometer we used had only 32 channels
per beam, and since the velocities with potential emission
were excluded from the baseline fits, the fits typically relied
on data from 22 to 25 channels only.  Given the small number of
channels available, we restricted the baseline fits to linear functions 
wherever possible.   In addition to the possibility of low-amplitude
standing waves across the bandpass of the observations, there
were also instances of individual channels that appeared to be
systematically high or low, again at a level that was subtle
enough not to be detected in an individual spectrum.  
We note that the errors shown in Figure \ref{panel} are the
statistical errors only.  The
systematic errors are much larger; we estimate that they are
at least comparable to and probably somewhat larger than the 
magnitude of the ratios for radii with $\rm r_{eff}
\ga$ 5 pc.   We conclude that the uncertainties in the 
average ratios measured over the five clouds measured at FCRAO
are probably $\pm$ 0.02 or higher.

\subsection{ NRAO 12 m Observations }

The OGC2 and Rosette clouds were observed with the NRAO 12 m telescope 
on Kitt Peak, AZ in 1993 June 15-20.  Typical system temperatures were 
350 -- 800 K at 115 GHz and 150 -- 400 K at 89 GHz and 98 GHz.  
For the Rosette observations,
the spectrometer was two 128 $\times$ 250 kHz filter banks which we
used to measure orthogonal polarizations; we used 
two 128 $\times$ 100 kHz filter banks for the OGC2 cloud.
At 100 GHz, the telescope half-power beam width is 63\arcsec.
We mapped each cloud in CO, HCN, and CS and tried to cover the 
entire area of each cloud; for the Rosette cloud, this meant
using a somewhat sparse sampling with a pointing spacing of 12\arcmin.
For the OGC2 cloud, we sampled at 1\arcmin\ spacing.

We removed linear baselines from the emission-free regions
of the spectra using the AT\&T Bell Labs package COMB.
For both clouds, we did not detect HCN or CS emission within
the sensitivity of our observations.  In the case of the
Rosette, Blitz \& Stark (1992, private communication) and 
Williams (1995) detected CS emission towards several
clumps; it is likely that we missed these clumps because of
our sparse sampling.  The 3$\sigma$ upper
limits to the ratios for these clouds are $\rm I_{HCN}/I_{CO}$
$\le$ 0.013 and $\rm I_{CS}/I_{CO}$ $\le$ 0.011 for the Rosette and
$\rm I_{HCN}/I_{CO}$ $\le$ .016 and $\rm I_{CS}/I_{CO}$ $\le$ 0.013
for OGC2.  The 3$\sigma$ upper limit for the Rosette ratio 
$\rm I_{CS}/I_{CO}$ is in good agreement with the ratio calculated
in Appendix B of Helfer \& Blitz (1993), $\rm I_{CS}/I_{CO}$ = 0.010.

\subsection{ Results of Cloud Study}

\begin{figure}
\vspace{-1.0in}
\plotone{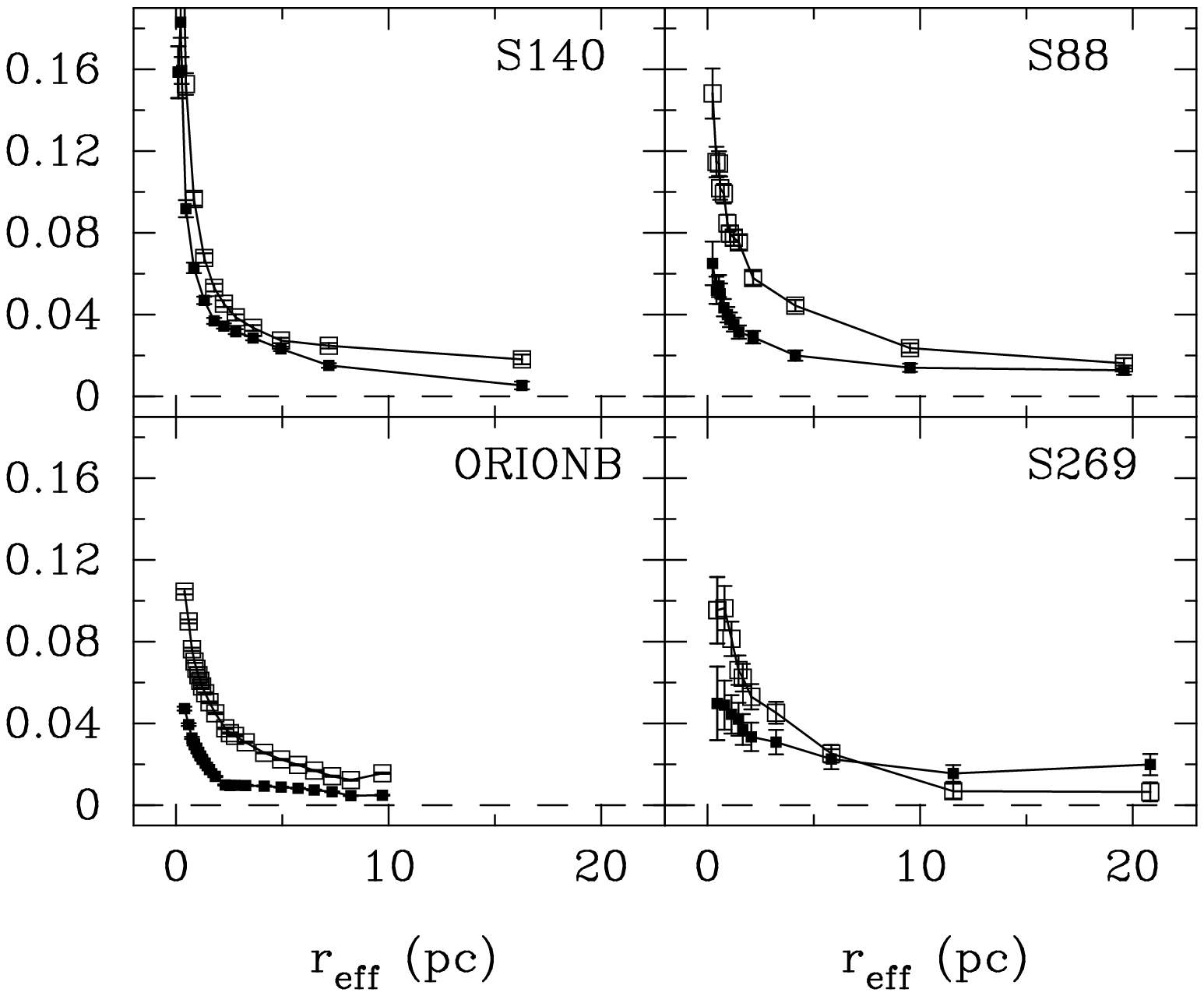}
\caption{The integral ratios $\rm I_{HCN}/I_{CO}$ (open squares) and 
$\rm I_{CS}/I_{CO}$ (filled squares) as a function of effective radius
in each of four GMCs measured with QUARRY.  The error bars underestimate 
the systematic 
uncertainties in the ratios at r$_{\rm eff}$ $\ga$ 5 pc (see text).}
\label{panel}\end{figure}

The cloud ratios for the FCRAO and NRAO studies are summarized
in Table 2\footnote{For the FCRAO data, we used the peaks
of the summed HCN and CS intensities in Figure \ref{fcraotot} to
compute the ratios.}.  If we take the 3$\sigma$ upper limits as the
measured ratios for the Rosette nondetections, then the average ratios
over the five GMCs measured are $<$$\rm I_{HCN}/I_{CO}$$>$ = 0.014 $\pm$ 0.02
and $<$$\rm I_{CS}/I_{CO}$$>$ = 0.013 $\pm$ 0.02.\footnote{Given the 
systematic problems with the FCRAO data, we assign both averages an 
uncertainty of $\pm$ 0.02, rather than the uncertainty in the
mean $\sigma$/N$^{0.5}$.}   These numbers are representative of
the average ratios over individual GMCs on size scales of
$\sim$ 40 pc diameter.  From Figure \ref{panel}, it is clear
that the star-forming cores that occur on size scales
of $\la$ 1 pc in GMCs are characterized by much higher ratios than 
is typical for most of the area of the GMC.  
This disparity is naturally explained by the difference
in the gas density between the cores and the ambient gas
across the extent of a GMC.

\begin{figure}
\plotfiddle{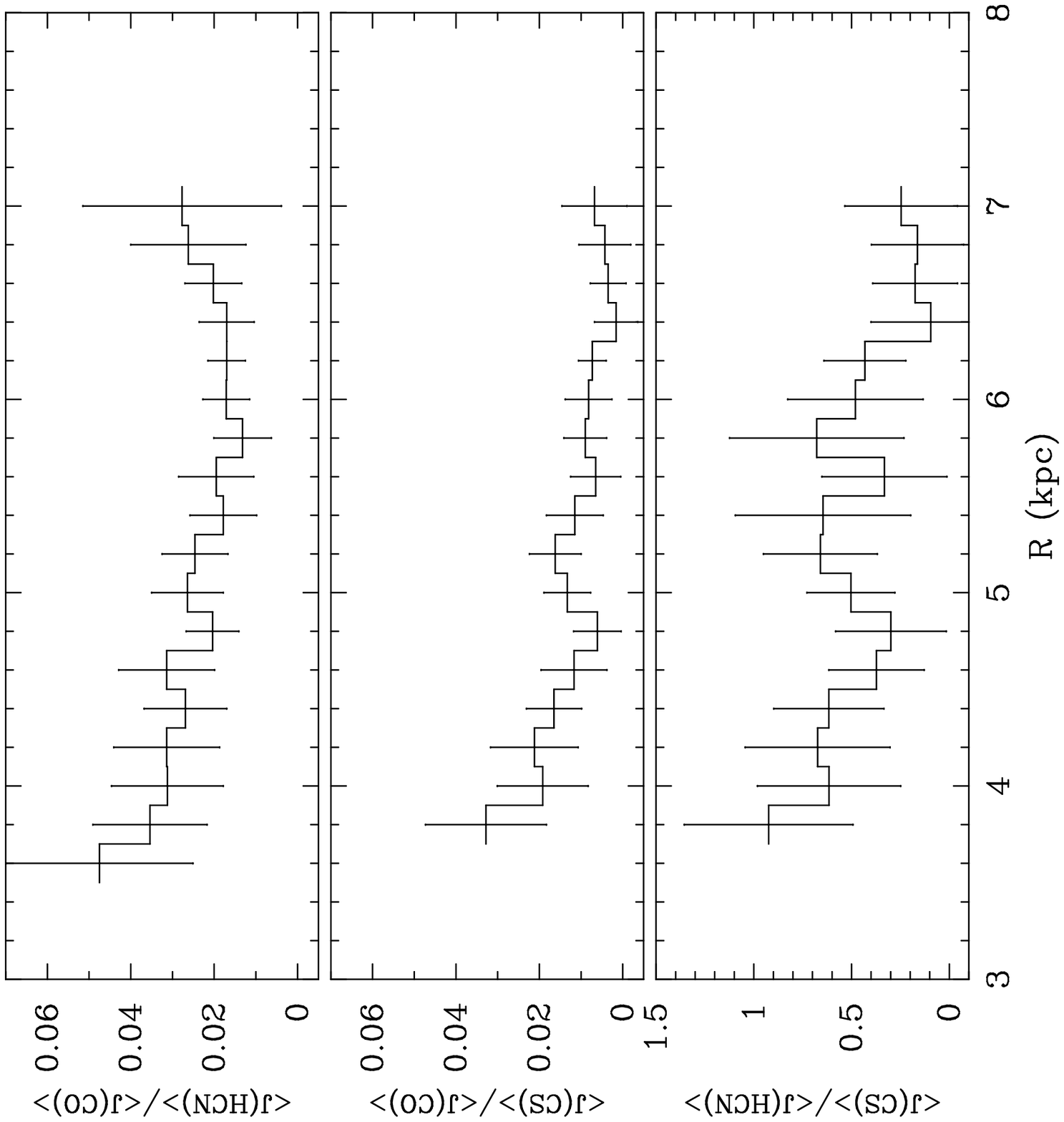}{400pt}{-90}{80}{80}{-288}{500}
\caption{The ratios of HCN, CS, and CO emissivities as a function of 
radius in the plane of the Milky Way. }
\label{planeratios}\end{figure}

\begin{figure}
\plotfiddle{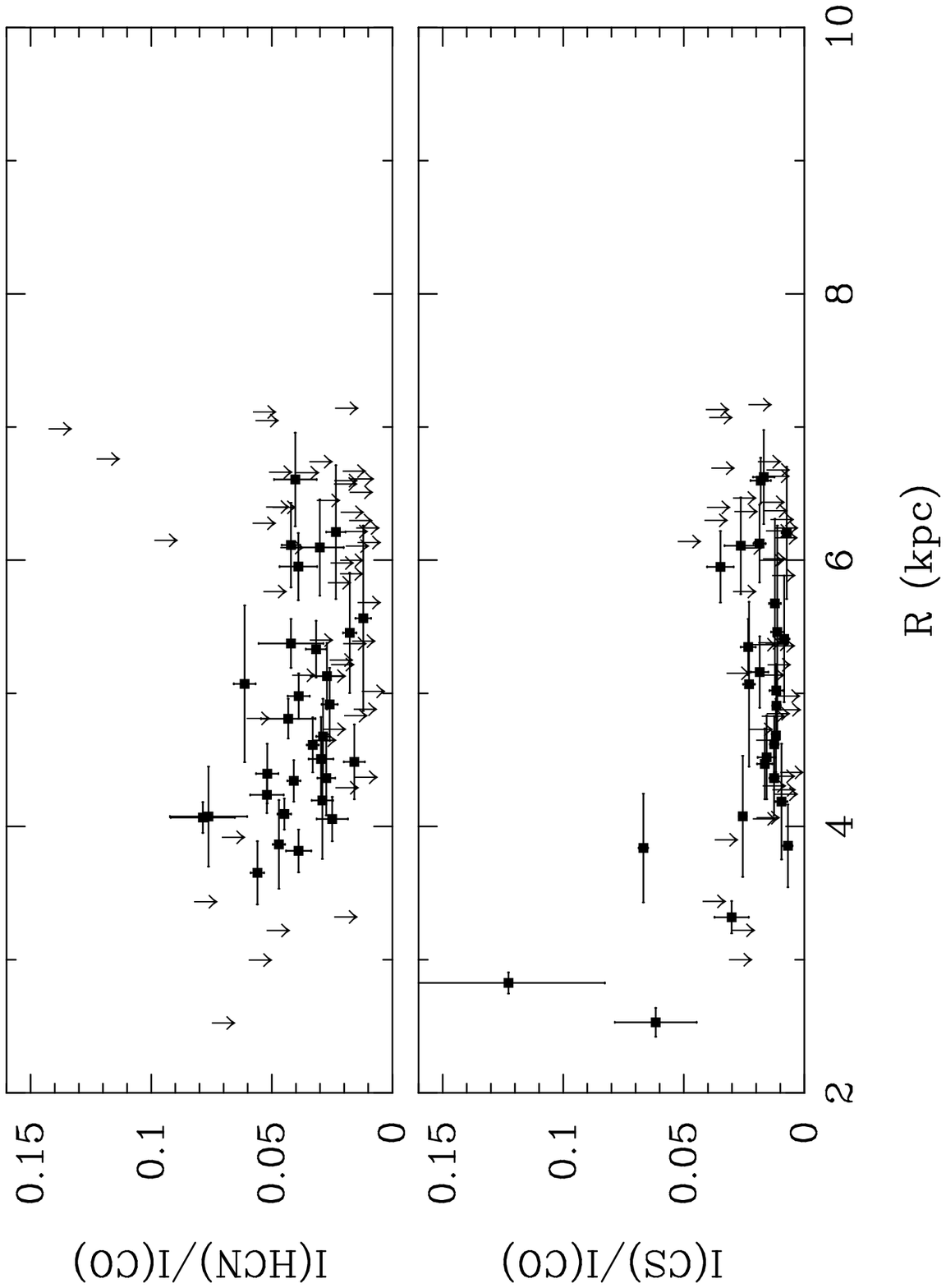}{400pt}{-90}{80}{80}{-288}{500}
\caption{The dense gas ratios over individual Milky Way features.  The
arrows represent 2$\sigma$ upper limits, and the filled squares
represent detected ratios.  The vertical error bars are the 1$\sigma$
uncertainties in the ratios, and the horizontal lines represent the
range of the Galactocentric radii implied by the velocity limits of each
feature. }
\label{features}\end{figure}

In cloud cores, the HCN emission usually dominates that of CS on 
scales of a few pc or smaller (see Figure \ref{panel}).  
This is also seen on all size scales ($\sim$ 1 pc - 3 kpc) in the 
Milky Way plane (see Figures \ref{planeratios} and \ref{features})
as well as over hundreds of pc in the centers of galaxies
(Helfer \& Blitz 1993; $\S$~5).
Over the entire GMCs, however, the ratios $\rm I_{HCN}/I_{CO}$ and
$\rm I_{CS}/I_{CO}$ are the same to within their measurement
errors.  Since the ratios over the entire GMCs are dominated
by their uncertainties, it is not clear if this effect is a
physical one or an artifact of the measurements.

\section{ Milky Way Plane Survey }

\subsection{ Observations}

We observed HCN, CS, and CO using an unbiased sampling in 
the first quadrant of the plane of
the Milky Way using the NRAO 12 m telescope in
1993 June 15 -- 20.  We made position-switched observations
using reference positions from Waller \& Tacconi-Garman (1992).
The system temperatures ranged from  300 -- 600 K at 115 GHz to
165 -- 500 K at 89 GHz and 98 GHz.  We observed orthogonal
polarizations and averaged the spectra from the two channels.
The backend was a hybrid spectrometer which was configured to
achieve 0.39 MHz (1.2 km s$^{-1}$ at 100 GHz) resolution over a 
300 MHz (900 km s$^{-1}$ at 100 GHz) bandwidth or 0.78 MHz resolution 
over a 600 MHz bandwidth .  Typical integrations were 2 minutes on
each position for the CO observations and 6 minutes for the
HCN and CS, though we were able to integrate longer on selected 
individual positions in HCN and CS.  The data were processed using the 
AT\&T Bell Laboratories package COMB.  Since there were standing waves 
in the 300 and 600 MHz bandpass, we fit polynomial functions to 
the baselines excluding channels in the velocity range where there 
was possible emission (-30 -- 140 km s$^{-1}$).  As a check to the fits, 
we also calculated a linear or quadratic fit to the baseline using 
emission-free channels over a limited range of the bandpass 
(-100 -- 250 km s$^{-1}$); the results were comparable to the
higher-order fits.

The survey coverage included positions in the Milky Way plane 
($b = 0\arcdeg$) at longitudes $15.\arcdeg5 \le l \le 55.\arcdeg5$
in steps of $\Delta l = 1 \arcdeg$.
Some of the spectra were rejected because of excessive standing
wave problems or other corruptions; in all, there were 41 positions
which were successfully observed in CO, 37 positions in HCN, and 31 
positions in CS.  There were 30 longitudes which were observed in all 
three species.  We use these 30 positions in the following analysis;
their spectra are shown in Figure \ref{planespectra}.

\subsection{ Geometric Considerations and Basic Results of Plane Survey}

For any given line of sight
$l$, an observed velocity $v_{\rm obs}$ corresponds to
a unique Galactocentric distance R via the geometric
relationship $v_{\rm obs}$ = V R/R$_{\rm o}$ sin $l$ - V$_{\rm o}$ sin $l$,
where V and V$_{\rm o}$ represent the circular velocities at distances
R and R$_{\rm o}$.  
The validity of this relationship breaks down for small longitudes
$l$, since large noncircular motions dominate in the region
of the Galactic Center.  We have avoided this complication by
restricting our observations to $l > 15$\arcdeg.  The 
relationship is also invalid for $|v_{\rm obs}| \le 20$ km s$^{-1}$
since the random motions of nearby clouds may dominate their
projected circular motions.

\begin{figure}
\plotfiddle{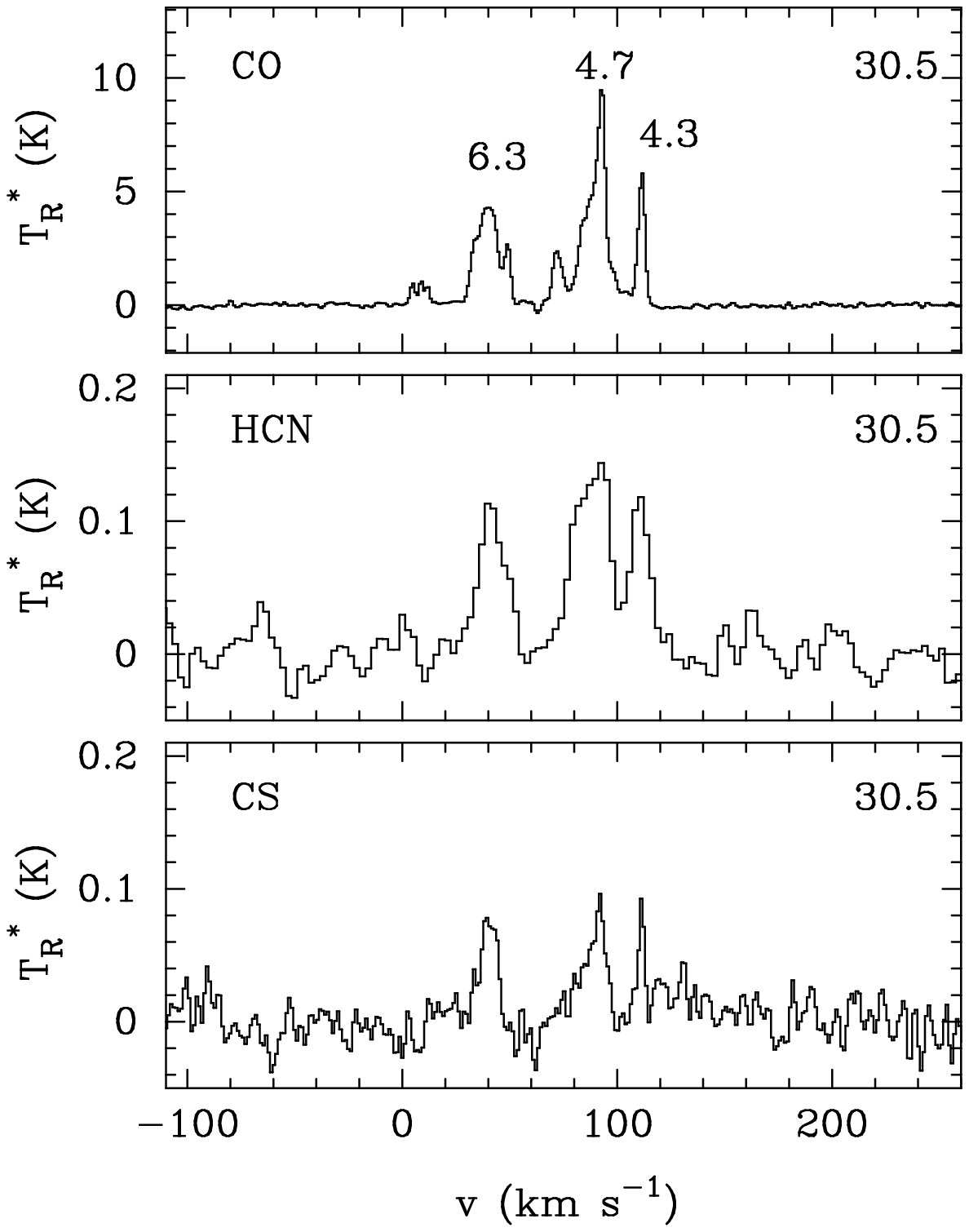}{400pt}{0}{60}{60}{-216}{-50}
\caption{CO, HCN, and CS spectra from $l$ = 30.5\arcdeg.  The three strongest
features in the CO spectrum are labeled with their Galactocentric distances
in kpc. }
\label{30degspectra}\end{figure}

We can illustrate some basic results of the plane survey by
examining the CO, HCN, and CS spectra at a sample longitude, 
$l$ = 30.5\arcdeg\ (Figure \ref{30degspectra}).
The CO spectrum at this longitude shows peaks at many different
velocities, which correpond to clouds at different 
distances\footnote{Throughout this study, we calculate kinematic
distances assuming a flot rotation curve with V$_{\rm o}$ =
220 km~s$^{-1}$ and R$_{\rm o}$ = 8.5 kpc.} along
the line of sight.  The strongest features in the CO spectrum,
those at 40 km s$^{-1}$, 90 km s$^{-1}$, and 110 km s$^{-1}$,
show corresponding emission in HCN and CS.   These features
are emitted from gas at kinematic distances from the Galactic
Center of 6.3 kpc, 4.7 kpc, and 4.3 kpc.  The feature at R = 6.3 kpc
could be at a heliocentric distance of 11.9 kpc or 2.7 kpc; at
these distances, the 1\arcmin\ resolution of the observations
corresponds to a linear resolution of 3.5 pc or 0.79 pc.
In general, the characteristic linear resolution
of the plane survey is $\sim$ 1 pc; this is much smaller than
the size scale of a typical GMC ($\sim$ 40 pc, see Figure
\ref{fcraomaps}), so a feature may be considered a 
``pencil-beam'' observation through some random location
in a GMC.

There are several points worth noting about the comparison of HCN and CS
features to the CO emission in Figure \ref{30degspectra}:  (1) The 
linewidths of the features
are the same (allowing for the triplet fine structure of the
HCN line, which causes a broadening of $\Delta v^{+5}_{-7}$ km s$^{-1}$).
(2) The shapes of the features are quite similar.  Both these
points emphasize the fact that the CO, the HCN and the CS emission 
at a given velocity all trace gas at the same physical region.
(3) The strengths of the HCN and CS emission vary with respect to
the CO emission from one feature to another; this suggests that
the excitation conditions or abundances vary among the different
positions.

\subsection{The Radial Distribution of Dense Gas Emission in the Plane}

In order to determine the dense gas distribution as a function
of radius in the plane of the Milky Way, we binned the inner
Galaxy in concentric annuli of width $\Delta$R = 200 pc and
used the CO, HCN, and CS spectra to calculate the emissivities
J$_l$ = 1/$\Delta$L $\int$T$_{\rm R}^*$ d$v$, where the velocity
limits of integration were determined for each bin from the 
relationship $v_{\rm obs}$ = V R/R$_{\rm o}$ sin $l$ - V$_{\rm o}$ sin $l$.
The emissivity represents the integrated intensity normalized by
the path length $\Delta$L through an annulus along the line
of sight to the Sun.  The data from all 30 longitudes were
then combined to form an average $<$J$>$(R) = 1/30 $\sum$$_{l}$ J$_l$(R)
for each of the three molecules.

\begin{figure}
\plotfiddle{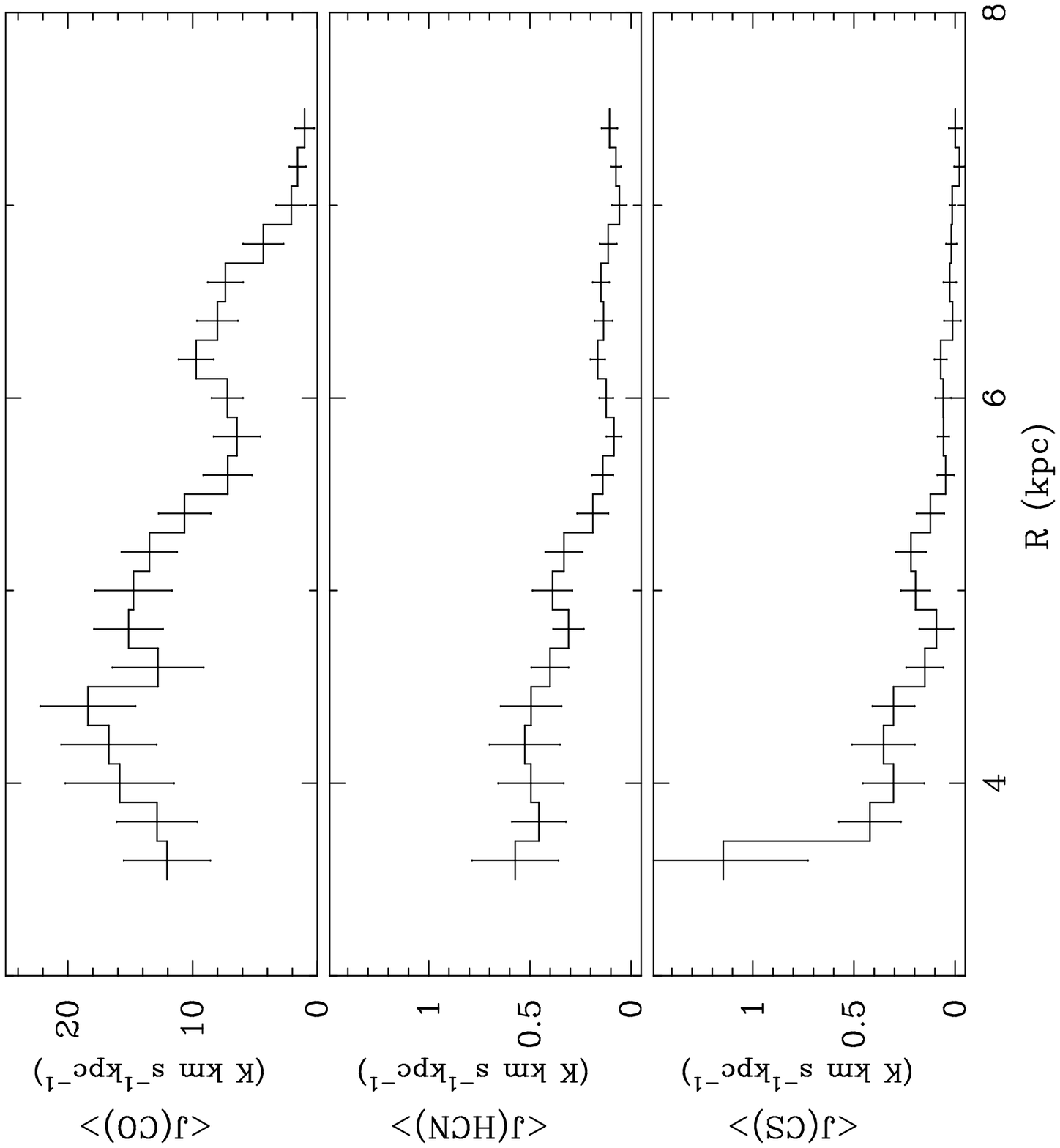}{400pt}{-90}{80}{80}{-288}{500}
\caption{Radial distributions of CO, HCN, and CS emissivities in the plane
of the Milky Way. }
\label{planeemission}\end{figure}

Figure \ref{planeemission} shows the radial distributions of CO,
HCN, and CS emissivities.  The CO distribution agrees well with
surveys of the Milky Way that include much more complete sampling
(e.g. Sanders, Solomon, \& Scoville 1984; Dame et al. 1987).  In particular,
there is a relative maximum of CO emission at radii in
the range 4 $<$ R $<$ 5.5 kpc that falls off with R at larger radii.
The emissivities from HCN and CS show similar distributions
to that of the CO, with an enhancement of emission at the
position of the relative maximum.

The ratios of emission are shown in Figure \ref{planeratios}.
There is a moderate tendency for higher $<$J$>$(HCN)/$<$J$>$(CO)
and $<$J$>$(CS)/$<$J$>$(CO) towards inner Galactocentric radii;
linear fits to the data yield $<$J$>$(HCN)/$<$J$>$(CO) = 0.049 
- 0.005 R and $<$J$>$(CS)/$<$J$>$(CO) = 0.040 - 0.005 R.  
The upturn in $<$J$>$(HCN)/$<$J$>$(CO) at R $\ga$ 6.6 kpc
suggests that local excitation effects might
have an important effect at a level of $\pm$ 0.01 in the ratios; 
alternatively, since the sensitivity of the survey is 
poorest at the endpoints, the values of the ratios may be artificially
high at these positions.  If the radial trend is real, we
note that it is modest in comparison with the typical 
measurement uncertainties.    The data are reasonably well 
represented by their average values (computed over ratios with
$> 2\sigma$ detections), $<$J$>$(HCN)/$<$J$>$(CO) = 
0.026 $\pm$ 0.008 and $<$J$>$(CS)/$<$J$>$(CO) = 0.018 $\pm$ 0.008 
over the region 3.5 $<$ R $<$ 7 kpc. 

The radial distribution of molecular cloud properties has also
been studied by Liszt, Burton, \& Xiang (1984), Liszt (1993), 
Sanders et al. (1993), Handa et al. (1993), Liszt (1995), and Sakamoto et
al. (1995, 1996);  the results of these studies are contradictory.
Liszt et al. (1984) studied the variation of the $^{12}$CO/$^{13}$CO
J=1--0 emission line ratio in the first quadrant of the Galactic
disk and found a tendency for lower $^{12}$CO/$^{13}$CO ratios towards inner
Galactocentric radii; they interpreted this result as evidence for
higher mean hydrogen column densities at inner radii.  However,
Liszt (1993) studied the J=2--1 to J=1--0 emission ratio of $^{13}$CO
in the same region and found no systematic variation across the
Galactic disk; he therefore concluded that the mean densities do not
vary as a function of radius and that abundance effects probably
cause the variation in the $^{12}$CO/$^{13}$CO ratios.  Liszt's
(1995) survey of HCO$^+$ in the first quadrant shows a
moderate tendency towards higher HCO$^+$/$^{13}$CO ratios at
inner radii, which he also interprets as abundance effects.
In contrast to Liszt's J=2--1 to J=1--0 $^{13}$CO study, 
Handa et al. (1993) and Sakamoto et al. (1995, 1996) presented an
extensive J=2--1 $^{12}$CO survey of the first quadrant, compared their 
data to the J=1--0 $^{12}$CO Columbia survey, and 
found a systematic gradient
across the disk; these authors concluded that the gas in the
inner galaxy is on average warmer and more dense than that
in the solar neighborhood.  Finally, Sanders et al. (1993)
used the J=3--2, J=2--1, and J=1--0 transitions of both
$^{12}$CO and $^{13}$CO and found no gradient across the disk 
in any of the line ratios. 
These contradictory studies as well as the work presented here all show that
if there is a trend in the physical properties of
disk clouds as a function of radius interior to the Solar circle,
then it is a modest one.

There is no clear radial trend in the ratio of HCN to CS
emissivity. On average, the HCN emissivity is stronger than the CS emissivity
by a factor of $\sim$ 2.  This is similar to what is observed in
the centers of other galaxies (Helfer \& Blitz 1993) 
and in the Milky Way (see below).   

\subsection{ Moderate Densities in the Plane ``Features''}

In nearby GMCs, less than 5\% of the area of a cloud has
detectable emission from HCN or CS at the sensitivity achievable
in an integration time of a few minutes (this study, see Figure
\ref{fcraomaps}).  The typical dense gas ratios 
over these pc-scale cores are $\rm I_{HCN}/I_{CO}$ $\ga$ 0.1.
We have seen that the ratios in the Milky Way plane are
much lower than this, or $\rm I_{HCN}/I_{CO}$ = 0.026 on 
average over size scales of 200 pc.  We may now ask how the average
ratios in the plane discussed in the previous section compare
with the ratios over individual, pc-scale ``features'' in 
the plane spectra of Figure \ref{planespectra}.  To do this, we
defined a ``feature'' by eye from an inspection of the CO
spectra at each longitude:  a ``feature'' was taken to
be either an isolated spectral line or, where the line blending
was too severe to choose a unique peak, the collection
of emission that made up the blended line.  Using limits
of integration that covered the CO emission, we recorded
the integrated intensity of CO and then measured the
HCN and CS emission over the corresponding velocities.
The ratios over these individual features are shown in
Figure \ref{features}.  

Although there is considerable
spread in the ratios, on average the values are low,
with the bulk of the ratios $\rm I_{HCN}/I_{CO}$ and
$\rm I_{CS}/I_{CO}$ between 0.01 -- 0.05.  These ratios are
consistent with the average ratios measured in the
plane, rather than the ratios measured in the dense
cores in GMCs (0.1 -- 0.3); this suggests that the plane features
are for the most part {\it not} dense cores that happen to intersect our line
of sight, but rather that the emission is from gas of
moderate density.  This conclusion is supported by
the low strengths of the HCN and CS lines in the plane:
although the kinetic temperature of the gas is on
the order of T$_{\rm K}$ = 10 -- 15 K (these temperatures
are consistent with the observed peak CO line strengths of T$_{\rm r}$
= 5 -- 10 K), the antenna temperatures of the HCN and CS
emission are typically 0.2 K or below (Figure \ref{planespectra}).
As the LVG modeling in $\S$ 2 suggests (see the lower middle panel
of Figure \ref{lvgfig}, which shows T$_{\rm r}$(HCN) as a function of
density for T$_{\rm K}$ = 15 K and N(H$_2$) $\sim$ 10$^{22}$ 
cm$^{-2}$), the densities of the plane features are probably only 
a few times 10$^3$ cm$^{-3}$.  

Liszt (1995) recently presented an unbiased survey of the 3 mm 
emission from HCO$^+$ (as well as selected positions in HCN, CS, 
and C$_2$H) in the Milky Way plane and reached a similar conclusion,
namely, that the ubiquity and low intensities of the emission
from these species imply that the gas is at moderate densities.

\section{ The Distribution of Dense Gas Ratios in the Milky Way }

To complete our synthesis of dense gas observations in the Milky
Way, we turn now to the center of the Galaxy, which was observed
in CS by Bally et al. (1987) and in HCN by Jackson et al. (1996).
In this region, CS and HCN emission are distributed throughout
the inner $d$ $\sim$ 500 pc ($-1\arcdeg \le l \le 1.\arcdeg8$),
which is characterized by the highest molecular emissivity in
the Galaxy (the surface density of CO is about 60 times 
higher in the bulge than the average at the Solar circle, Sanders 
et al. 1984).  The dense gas emission extends some 12 pc
(0.08\arcdeg) out of the plane of the Milky Way (Jackson et al. 1996).
Averaged over the central 500 pc, the CS to CO ratio is 
$\rm I_{CS}/I_{CO}$ = 0.027 $\pm$ 0.006 (Helfer \& Blitz 1993).
The HCN to CO ratio ranges from 0.04 to about 0.12 on small scales within the
bulge region; averaged over the central 600 pc, the ratio is 
$\rm I_{HCN}/I_{CO}$ = 0.081\footnote{
Jackson et al. 1996.  We use their correction for HCN
and CO emission at nonzero Galactic latitude.  In their ``aperture 
photometry'', Jackson et al. smooth their HCN map to a spatial
resolution larger than that of their CO map by a factor of 
$\rm \lambda(HCN)/\lambda(CO)$ in order to emulate the single-dish 
measurements of extragalactic ratios.  This correction is not appropriate
for comparison with this study; we therefore ``re-correct'' their ratio 
so that it is appropriate for direct comparison to these results. } 
$\pm$ 0.004.  We note that the errors quoted here are formal uncertainties;
these may underestimate the uncertainties due to absolute calibration
errors.

\begin{figure}
\plotfiddle{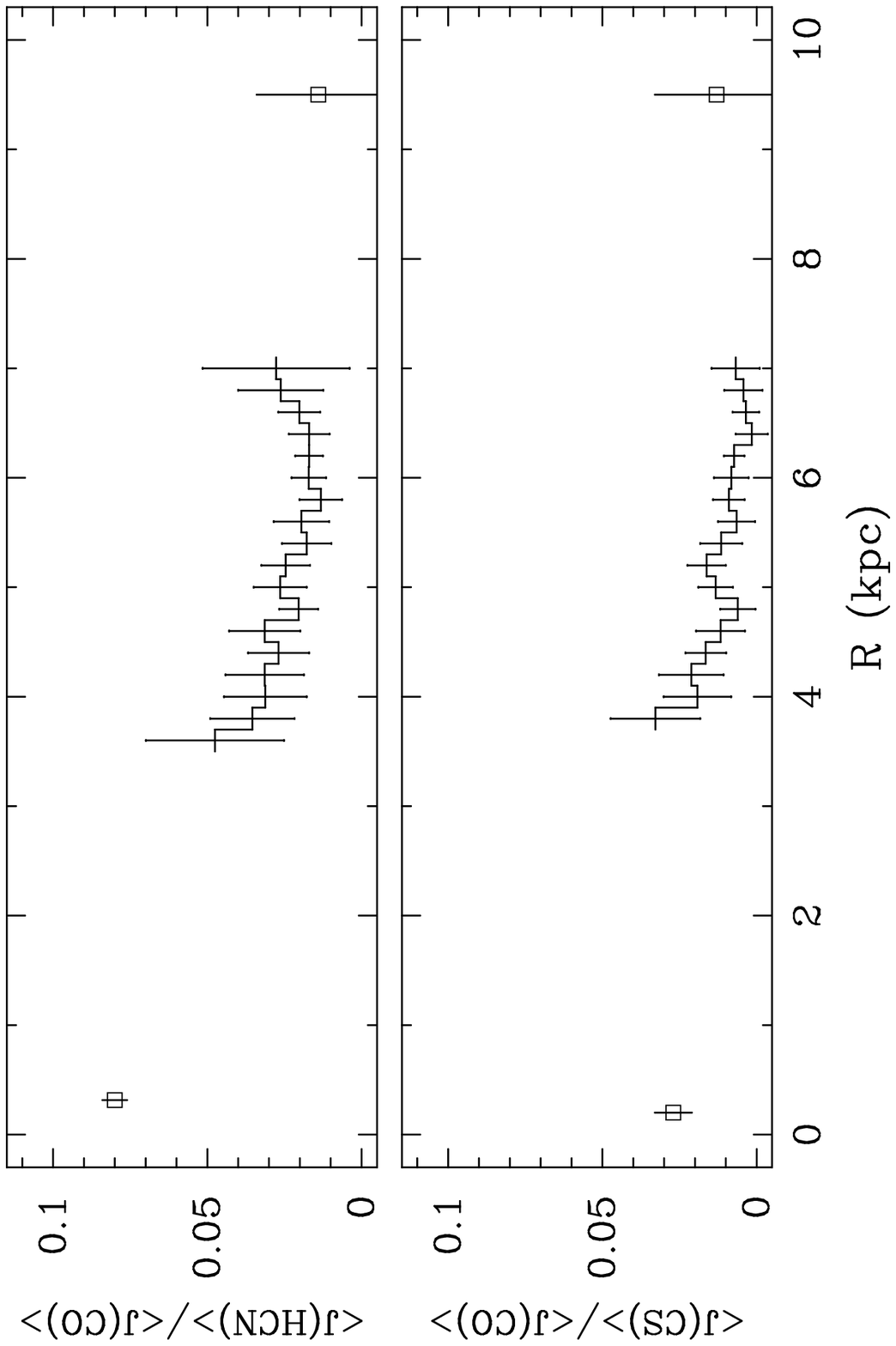}{400pt}{-90}{80}{80}{-288}{400}
\caption{Dense gas ratios in the Milky Way.  At R = 0.3 kpc, the HCN/CO point 
is from Jackson et al. 1996, and the CS/CO point is from Helfer \& Blitz
1993.  The points shown at R = 9.5 kpc represent the averages over the
individual GMCs presented in this study; the clouds themselves are located
at different directions and distances from the Sun. } 
\label{mwratios}\end{figure}

Armed with the HCN and CS studies of the Galactic Center (Bally et al. 1987;
Helfer \& Blitz 1993; Jackson et al. 1996), the plane of the Milky Way 
($\S$ 4), and
clouds near the Sun ($\S$ 3), we now wish to compare the
distribution of the ratios of dense gas as a function of location
in the Milky Way.  These ratios are summarized in Figure \ref{mwratios}.
Clearly, there is a radial trend in the ratios:
they are higher by a factor of 3 in $\rm I_{HCN}/I_{CO}$
and 1.5 in $\rm I_{CS}/I_{CO}$ in the center of the Milky Way when
compared with the average ratios in the disk.
Although the plane $\rm I_{HCN}/I_{CO}$ ratio is about 0.01
higher than that measured over nearby GMCs (the $\rm I_{CS}/I_{CO}$ differs
by even less), the differences are within the measurement uncertainties.
We therefore conclude that, while there might be a modest trend
towards higher ratios for smaller radii for R $>$ 3 kpc, by
far the dominant result is the contrast between the high ratios
in the Milky Way bulge and the lower ratios in the disk.
The high ratios in the bulge are most likely caused by a combination
of higher gas densities as well as higher kinetic temperatures
in the bulge gas: n(H$_2$) $\sim$ 10$^4$ cm$^{-3}$ and T$_{\rm K}$ 
$\sim$ 70 K in the bulge, while n(H$_2$) $\sim$ 10$^{2.5-3}$ cm$^{-3}$
and T$_{\rm K}$ $\sim$ 15 K in the disk (G\"{u}sten 1989 and references
therein).
The correlation between the dense gas ratios and the known 
properties of molecular gas in the bulge and disk of the Milky Way
gives credibility to the use of dense gas to CO emission ratios as a
tracer of different physical properties in galaxies.

\section{The Relation Between Dense Gas Ratios and Pressure }

Any diffuse gaseous component in the Milky Way has a total 
internal pressure that is at least as high as that determined 
by the condition of hydrostatic equilibrium with the stellar 
gravitational potential.  The total gas pressure may be
represented as $\rm P_g = 2 \pi G \rho_* \Sigma_g h_g$, 
where $\rho_*$ is the stellar density, $\rm \Sigma_g$ is the total
gas surface density, and $\rm h_g$ is the scale height of the gas
(Spergel \& Blitz 1992).   Using this expression for the
total gas pressure, we calculated the pressure at the positions of 
five points for which we have measured values for
the ratio $\rm I_{HCN}/I_{CO}$:  the Galactic Center, three
positions in the Milky Way plane (using binned data from Figure
\ref{planeratios}), and the average ratio over individual
GMCs.  For $\rho_*$, we assumed an
exponential distribution as a function of radius, with
a scale length of 3 kpc (Spergel, Malhotra, \& Blitz 1996);  
to set the scale for $\rho_*$, we used the local value
at the solar circle of 0.1 $\rm M_{\sun} pc^{-3}$ (Gould 1995).
We used the CO and HI surface densities compiled by Dame (1993).
The CO scale heights
$\rm h_g$ are from Sanders et al. (1984) and Dame et al. (1987). At 
the Galactic Center, we used $\rho_*$ = 25 $\rm M_{\sun} pc^{-3}$ (Spergel
\& Blitz 1992) and $\rm \Sigma_g$(CO) = 180 $\rm M_{\sun} pc^{-2}$
(based on Sanders et al. 1984).

\begin{figure}
\plotfiddle{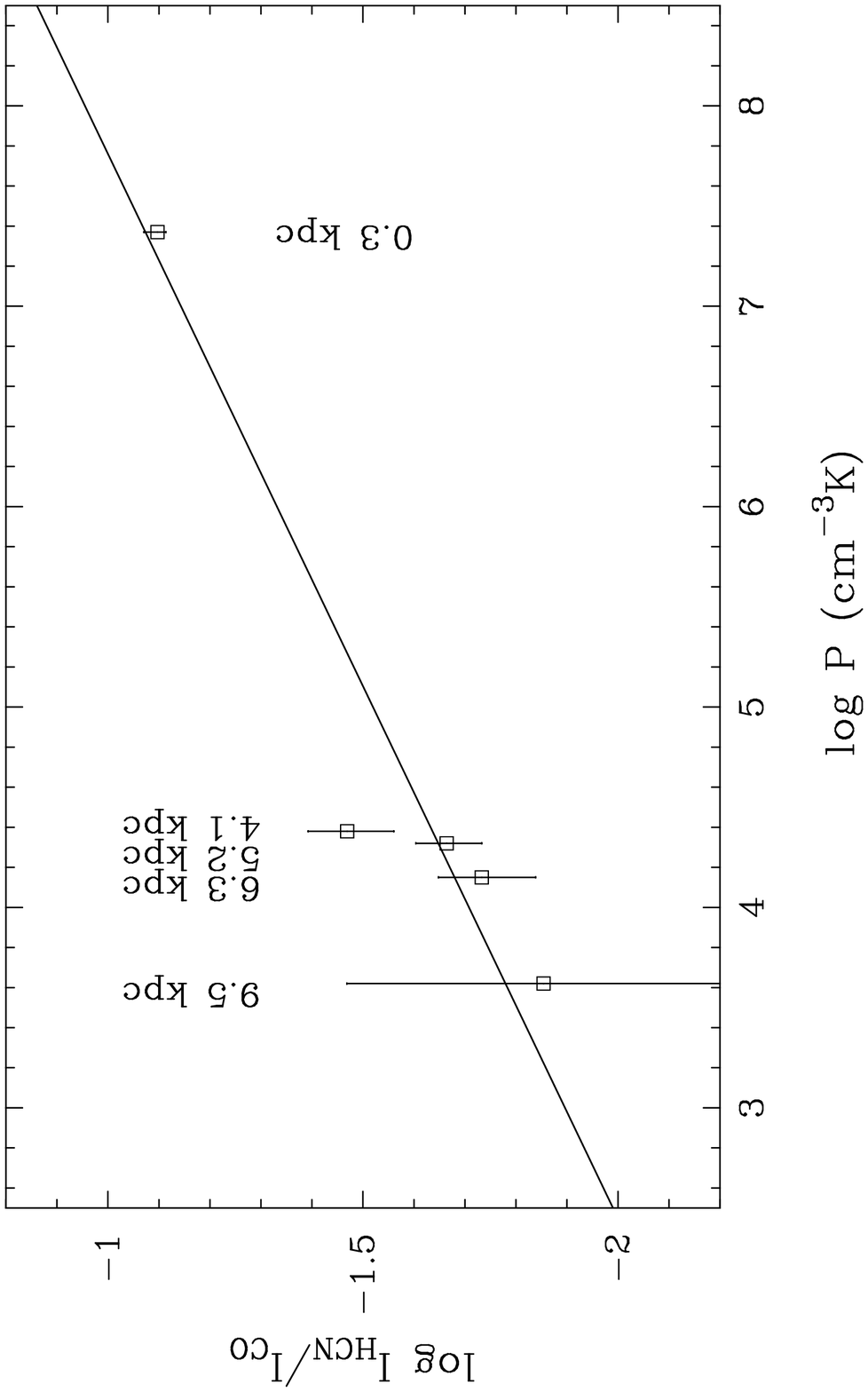}{460pt}{-90}{60}{60}{-230}{500}
\caption[Dense gas ratios vs. pressure in the Milky Way]{Dense gas ratios 
plotted against pressure in the Milky Way.  Each point is labeled with
its distance from the Galactic Center. The solid line is a linear least 
squares fit to the data. } 
\label{pressure}\end{figure}

The results of this calculation are shown in Figure \ref{pressure}.
Each point is labeled with its distance from the Galactic Center,
so the progression in both pressure and in $\rm I_{HCN}/I_{CO}$
is correlated with distance.  A fit to the data yields the
result that the ratios $\rm I_{HCN}/I_{CO}$ rise with pressure 
as $\rm I_{HCN}/I_{CO}$ $\propto$ P$^{0.19\pm0.04}$. 
In $\S$ 2, the claim was made that
the ratio of emissions $\rm I_{HCN}/I_{CO}$ could be used
as a qualitative indicator of pressure in a galaxy.  Figure
\ref{pressure} suggests that the ratio $\rm I_{HCN}/I_{CO}$
may be a reasonable {\it quantitative} measure of pressure.
It remains a topic of future research to determine whether this
relation holds in external galaxies, and if so, whether the
slope of the correlation is a universal one.

\section{Conclusions}

We have presented a systematic examination of the 3 mm emission
from HCN, CS, and CO on size scales of GMCs ($\sim$40 pc) and
larger in the bulge and disk of the Milky Way.  This study
combined new observations of individual GMCs and the Milky
Way plane with published studies of the inner 500 pc of the 
Galaxy.

\subsection{ Results from the GMC survey}

We surveyed five molecular clouds with the FCRAO 14 m telescope
using the QUARRY multibeam receiver and an additional two
clouds using the NRAO 12 m telescope.  As seen in the FCRAO
maps, less than 5\% of the area of a cloud has detectable
emission from HCN or CS at the sensitivity achievable in an
integration time of a few minutes.  The ratios $\rm I_{HCN}/I_{CO}$
and $\rm I_{CS}/I_{CO}$ are strong functions of the effective
radius of the clouds: the peak values for $\rm I_{HCN}/I_{CO}$ range 
from 0.1 -- 0.3 over size scales of $\le$ 1 pc, but by r$_{\rm eff}$
= 5 pc, $\rm I_{HCN}/I_{CO}$ drops to 0.04 or below, and by
r$_{\rm eff}$ $\ge$ 10 pc, $\rm I_{HCN}/I_{CO}$ is $<$ .02.
Similar trends are seen for $\rm I_{CS}/I_{CO}$.  Taking into
account the systematic uncertainties in the data, the
average ratios over the GMCs measured are $<$$\rm I_{HCN}/I_{CO}$$>$
= 0.014 $\pm$ 0.02 and $<$$\rm I_{CS}/I_{CO}$$>$ = 0.013 $\pm$
0.02.  These numbers represent the average ratios over individual
GMCs near the Sun on size scales of $\sim$ 40 pc diameter.
The disparity between the higher ratios in the pc-sized cloud cores
and the low ratios characteristic of entire GMCs is naturally
explained by the difference in gas density between the cores and
the ambient gas across the full extent of a GMC.

\subsection{ Results from the Milky Way Plane Survey}

We surveyed HCN, CS, and CO using an unbiased sampling over
some 40\arcdeg\ of the first quadrant of the plane of the Milky 
Way using the NRAO 12 m telescope.  Features from HCN and CS
were surprisingly common.  When compared to the corresponding
CO emission, the HCN and CS features showed the same linewidths
and line shapes, which implies that the CO, HCN, and CS emission
at a given velocity all trace gas from the same physical region.
However, the relative strengths of the lines vary from one
feature to another; this suggests that the excitation conditions
or abundances vary among the different positions.

On average, the ratios $\rm I_{HCN}/I_{CO}$ and $\rm I_{CS}/I_{CO}$
range between 0.01 -- 0.05 over individual, pc-sized features
in the plane survey.  These ratios are much lower than those
measured in the cores of GMCs (0.1 -- 0.3); this suggests that
the features are for the most part {\it not} dense cores that
happen to intersect our line of sight, but rather that the
gas is at moderate density.  This conclusion is supported by the
relatively weak strengths of the emission lines seen in HCN and CS.
These results were also noted in the HCO$^+$ survey of the
Milky Way plane by Liszt (1995).  

We binned the plane data as a function of Galactocentric radius
in concentric annuli of width 200 pc and computed the
average $\rm I_{HCN}/I_{CO}$ and $\rm I_{CS}/I_{CO}$ for the
different bins.   There is a moderate tendency for higher
ratios towards inner Galactocentric radii; however, the 
data are well represented by their averages: $\rm I_{HCN}/I_{CO}$ = 
0.026 $\pm$ 0.008 and $\rm I_{CS}/I_{CO}$ = 0.018 $\pm$ 0.008 over
the region 3.5 $<$ R $<$ 7 kpc.

\subsection{ General Results}

We find a strong trend in the ratios of HCN and CS emission 
as a function of location in the Milky Way:  in the bulge,
$\rm I_{HCN}/I_{CO}$ = 0.081 $\pm$ 0.004, in the plane,
$\rm I_{HCN}/I_{CO}$ = 0.026 $\pm$ 0.008 on average, and
over the full extent of nearby GMCs, $\rm I_{HCN}/I_{CO}$
= 0.014 $\pm$ 0.02.  Similar trends are seen for CS:
in the bulge, $\rm I_{CS}/I_{CO}$ = 0.027 $\pm$ 0.006,
in the plane, $\rm I_{CS}/I_{CO}$ = 0.018 $\pm$ 0.008 on
average, and over the full extent of nearby GMCs, 
$\rm I_{CS}/I_{CO}$ = 0.013 $\pm$ 0.02.  (Formal
uncertainties are quoted; these may underestimate the
uncertainties due to absolute calibration errors.)

The dominant result of this study is the contrast between
the high ratios in the Milky Way bulge and the lower
ratios in the disk.  The high bulge ratios are likely
caused by a combination of higher gas densities and
higher kinetic temperatures in the bulge gas (G\"usten
1989).  The correlation between the HCN and CS emission ratios
and the known properties of molecular gas in the bulge and 
disk of the Milky Way gives credibility to the use of
dense gas to CO emission ratios as a tracer of different
physical properties in galaxies.

Using stellar and gas surface densities from the literature,
we calculated the total gas pressure as a function of
radius in the Milky Way and compared the results with
the $\rm I_{HCN}/I_{CO}$ ratios we measured.  A
fit to the data shows that the ratios $\rm I_{HCN}/I_{CO}$
rise with pressure as $\rm I_{HCN}/I_{CO}$ $\propto$ P$^{0.19\pm0.04}$.
In this paper, we have claimed that $\rm I_{HCN}/I_{CO}$ could
be used as a qualitative indicator of pressure in a galaxy;
this result suggests that $\rm I_{HCN}/I_{CO}$ may be a
reasonable quantitative measure of pressure.

\acknowledgements

We thank the referee, Jim Jackson, for useful comments on the
manuscript.
We thank Mark Heyer for his advice with the FCRAO observations
and for supplying us with with CO map of S140; thanks also
to the FCRAO staff for making the cloud observations possible.  
We thank the staff at NRAO for their assistance with the plane 
and cloud observations.
TTH thanks Jack Welch for his hospitality while visiting UC Berkeley,
and she thanks John Lugten for many discussions about the details
of the data analysis and the resulting science.  This research
was supported by grants from the National Science Foundation
and the State of Maryland.

\clearpage


\begin{deluxetable}{lrrrrr}
\tablewidth{5.8truein}
\tablecaption{Cloud Sources }
\tablehead{
\colhead{Cloud}                 &
\colhead{$l$}       		&
\colhead{$b$}       		&
\colhead{$d$}			&
\colhead{$\rm R$}		&
\colhead{$v$$_{\rm LSR}$\tablenotemark{a}}      \\[.2ex]
\colhead{}			&
\colhead{(\arcdeg)}		&
\colhead{(\arcdeg)}		&
\colhead{(kpc)}			&
\colhead{(kpc)}			&
\colhead{(km~s$^{-1}$)}		
}
 
\startdata
S88                     & 61.47       &  0.10   & 2.0   & 7.8  & 21   \nl
S140                    & 106.76      &  5.28   & 0.9 & 8.8  & -8   \nl
S269                    & 196.45      &  -1.68  & 3.8 & 12.2 & 17.5 \nl
Orion B                 & 206.53      & -16.35  & 0.4 & 8.6  & 10   \nl
Rosette                 & 206.86      &  -2.21  & 1.6 & 10.0 & 16   \nl
			&	      & 	&     &	     &      \nl
OGC2\tablenotemark{b}   & 137.75      &  -1.00  & 21  & 28   & -103  \nl
MBM 32\tablenotemark{c} & 147.20      &  40.67  & 0.1 & 8.6  & 0    \nl
\enddata
\tablenotetext{a}{ $v$$_{\rm LSR}$ is defined by the radio convention, 
$v_{\rm LSR}$/c = $\Delta$$\lambda$/$\lambda$$_o$, where $\lambda$$_o$ 
is the wavelength in the rest frame of the source.}
\tablenotetext{b}{ Outer Galaxy Cloud 2, from Digel, de Geus, \& Thaddeus 1994.}
\tablenotetext{c}{ High-latitude cloud from Magnani, Blitz, \& Mundy 1985.}
\end{deluxetable}


\begin{deluxetable}{lrrr}
\tablewidth{5.8truein}
\tablecaption{Results from Cloud Surveys }
\tablehead{
\colhead{Cloud}                 &
\colhead{$\rm I_{HCN}/I_{CO}$}   &
\colhead{$\rm I_{CS}/I_{CO}$}    &
\colhead{Telescope}
}
 
\startdata
S88       & 0.016\tablenotemark{a}	& 0.013  		& FCRAO \nl
S140      & 0.019			& 0.013  		& FCRAO \nl
S269      & 0.007\tablenotemark{b}      & 0.020  		& FCRAO \nl
Orion B   & 0.014		        & 0.006\tablenotemark{b,c}& FCRAO \nl
Rosette   & $\le$ 0.013\tablenotemark{d}& $\le$ 0.011\tablenotemark{c}& NRAO\nl
	  &				&			& 	\nl
OGC2      & $\le$ 0.016 		& $\le$ 0.013  		& NRAO  \nl
MBM 32    & $\le$ 0.015 		& $\le$ 0.009   	& FCRAO \nl
	  &				&			& 	\nl
GMC Averages  & 0.014 $\pm$ 0.020\tablenotemark{e}& 0.013 $\pm$ 0.020\tablenotemark{e}  &  \nl
\enddata
\tablenotetext{a}{Uncertainties are $\pm$ 0.02 for all FCRAO GMC data, see 
$\S$ 3.1.3.} 
\tablenotetext{b}{The S269 HCN data and the Orion B CS data showed clear
evidence for the systematic problems discussed in $\S$ 3.1.3.  }
\tablenotetext{c}{These results compare well with those in Appendix B of
Helfer \& Blitz 1993, where we calculated $\rm I_{CS}/I_{CO}$ = 0.008 
for Orion B and $\rm I_{CS}/I_{CO}$ = 0.010 for the Rosette.}
\tablenotetext{d}{3$\sigma$ upper limits.} 
\tablenotetext{e}{See $\S$ 3.1.3 and footnote 7 for a discussion 
of the uncertainties.}
\end{deluxetable}

\clearpage

\begin{figure}
\plotfiddle{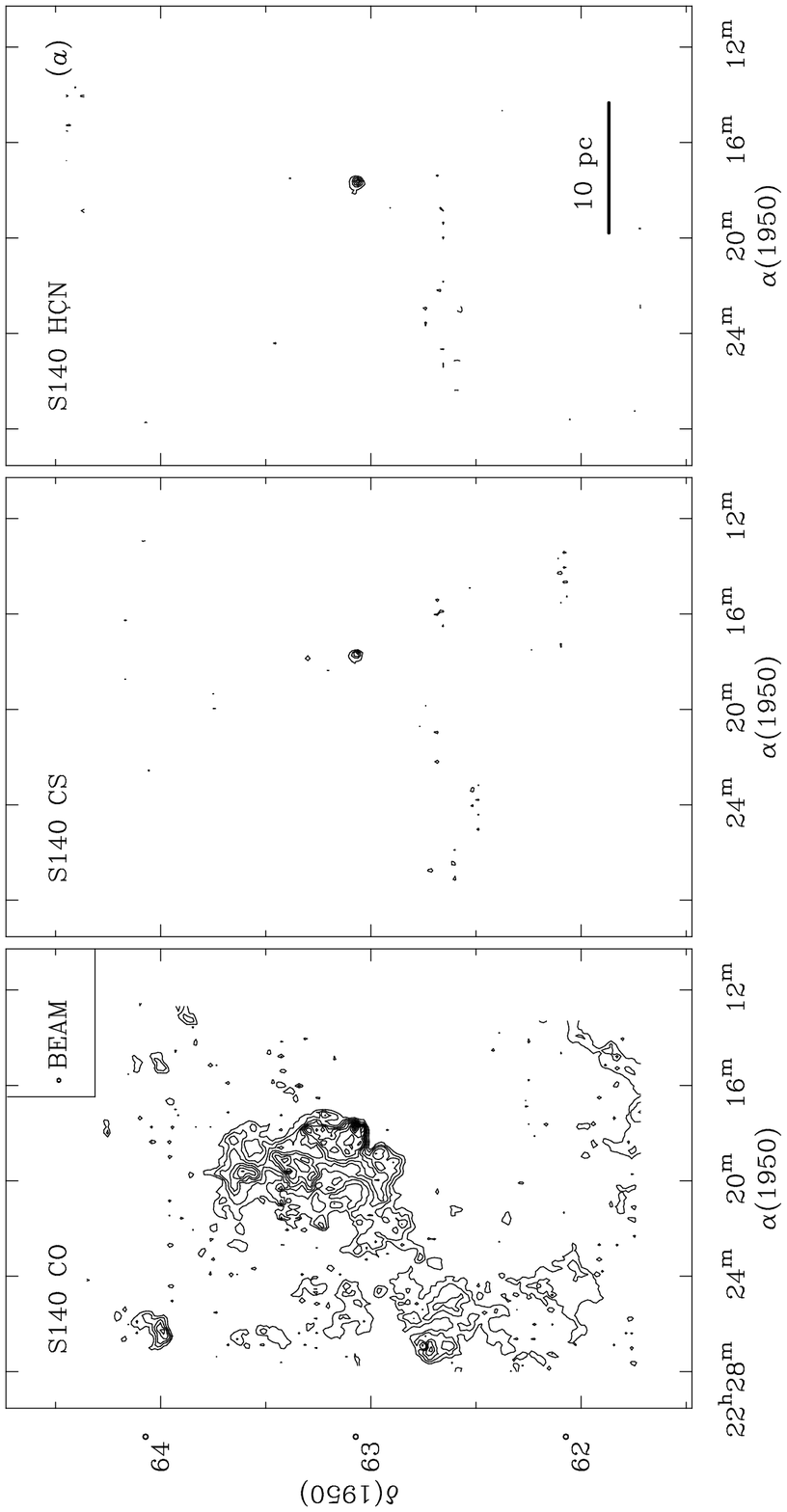}{360pt}{-90}{70}{70}{-255}{550}
\vspace*{-1in}
\plotfiddle{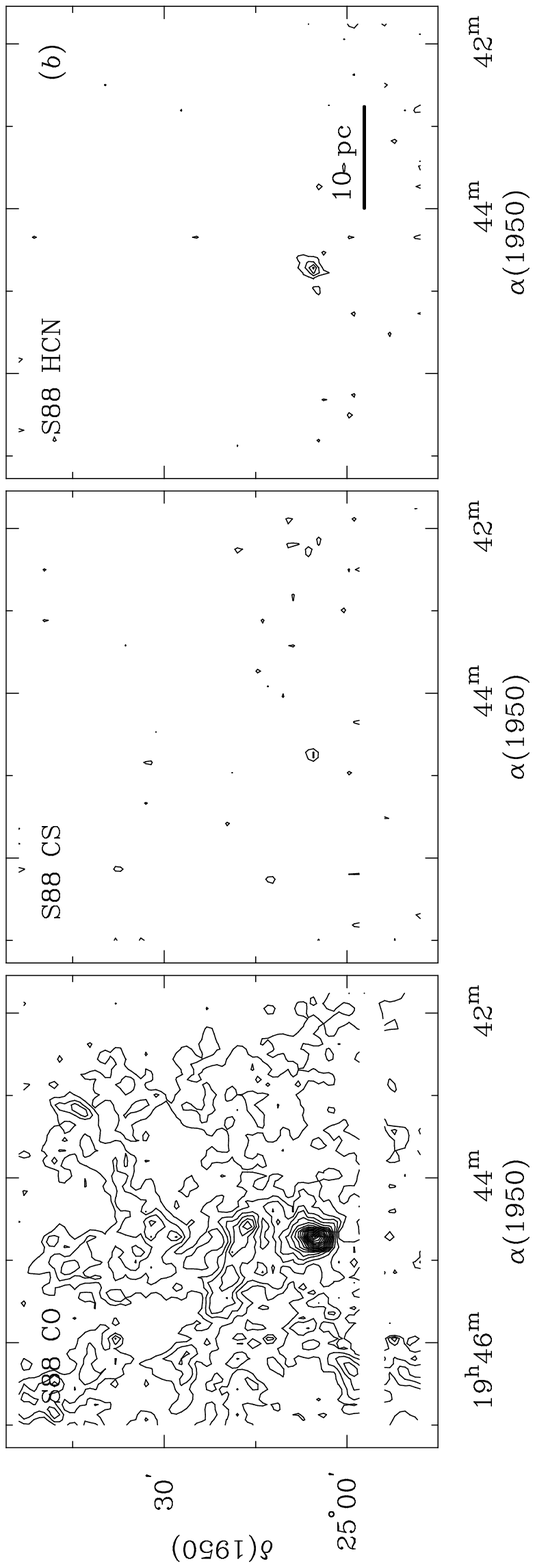}{-180pt}{-90}{70}{70}{-255}{300}
\caption{($a$) CO, CS, and HCN emission from the GMC S140.  The maps were
made using the QUARRY multibeam receiver at the FCRAO 14 m telescope.
Each map comprises some 23000 spectra.  The 1\arcmin\ resolution of 
the observations is shown in the upper right corner of the CO map, and
the characteristic size scale of 10 pc is shown on the HCN map.
The CO contour levels are $\pm$ 5,10,15... K km s$^{-1}$, and the
CS and HCN contours are $\pm$ 3,6,9... K km s$^{-1}$.  The
3 K km s$^{-1}$ contours some $\sim$ 15\arcmin\ to the north of the
peak on the CS and HCN maps show real emission in their spectra; all of
the 3 K km s$^{-1}$ contours farther than $\sim$ 15\arcmin\ away from 
the peak position on the CS and HCN maps are consistent with noise. The 
CO map is courtesy of Mark Heyer. ($b$) Same for S88.  The 
CO and CS contours are as above; 
the HCN contours are $\pm$ 3.3,6.6,9.9,... K km s$^{-1}$.}
\end{figure}

\begin{figure}
\plotfiddle{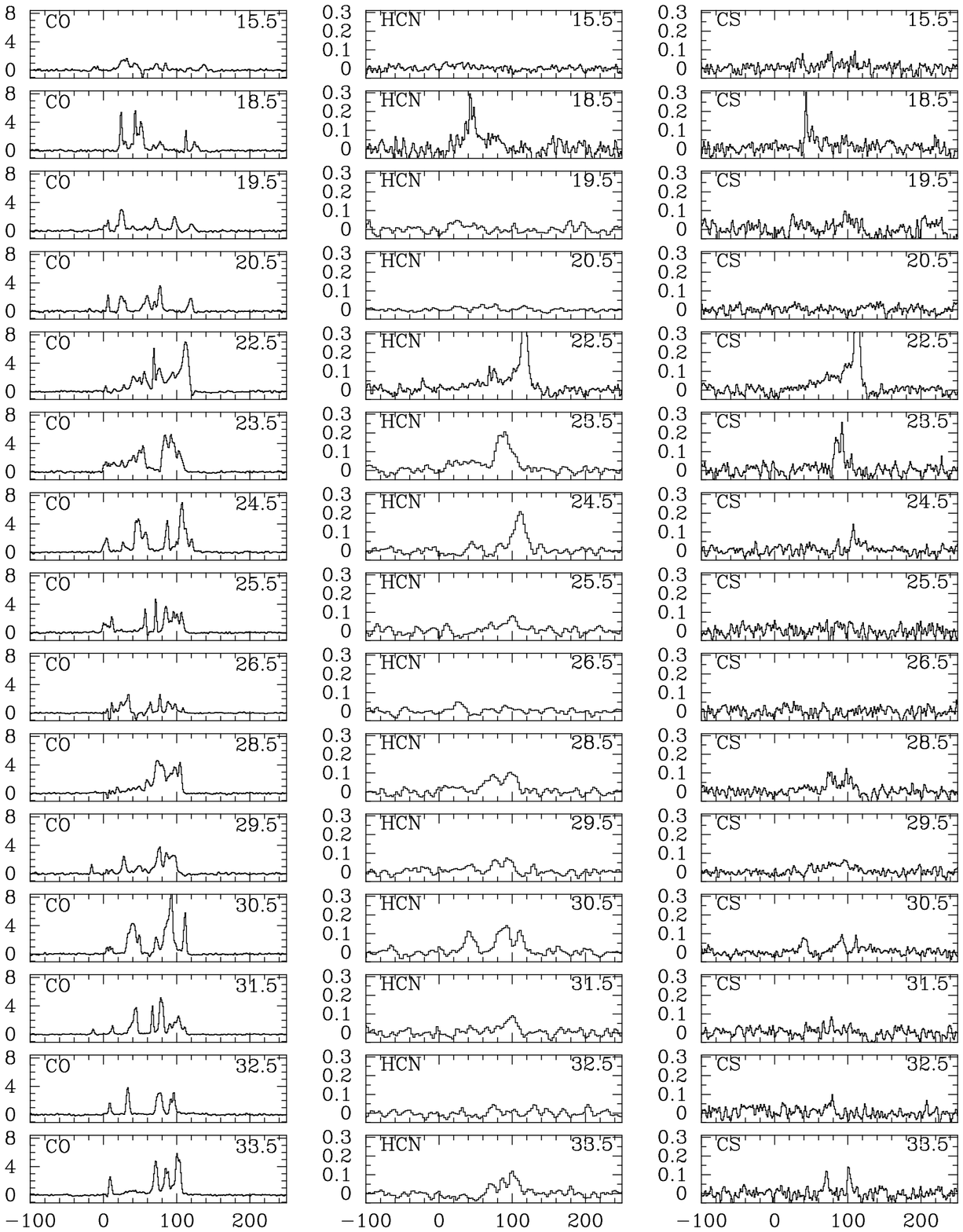}{460pt}{0}{80}{80}{-248}{-72}
\caption{CO, HCN, and CS spectra from NRAO 12 m plane survey.  The
longitude is given in the upper right corner of each spectrum.  The
abscissa is v$_{\rm LSR}$ in km s$^{-1}$, and the ordinate is 
T$_{\rm R}^*$ in K.  Note that the HCN and CS spectra have a
different temperature scale than the CO spectra.}
\label{planespectra}\end{figure}

\begin{figure}
\figurenum{5}
\plotfiddle{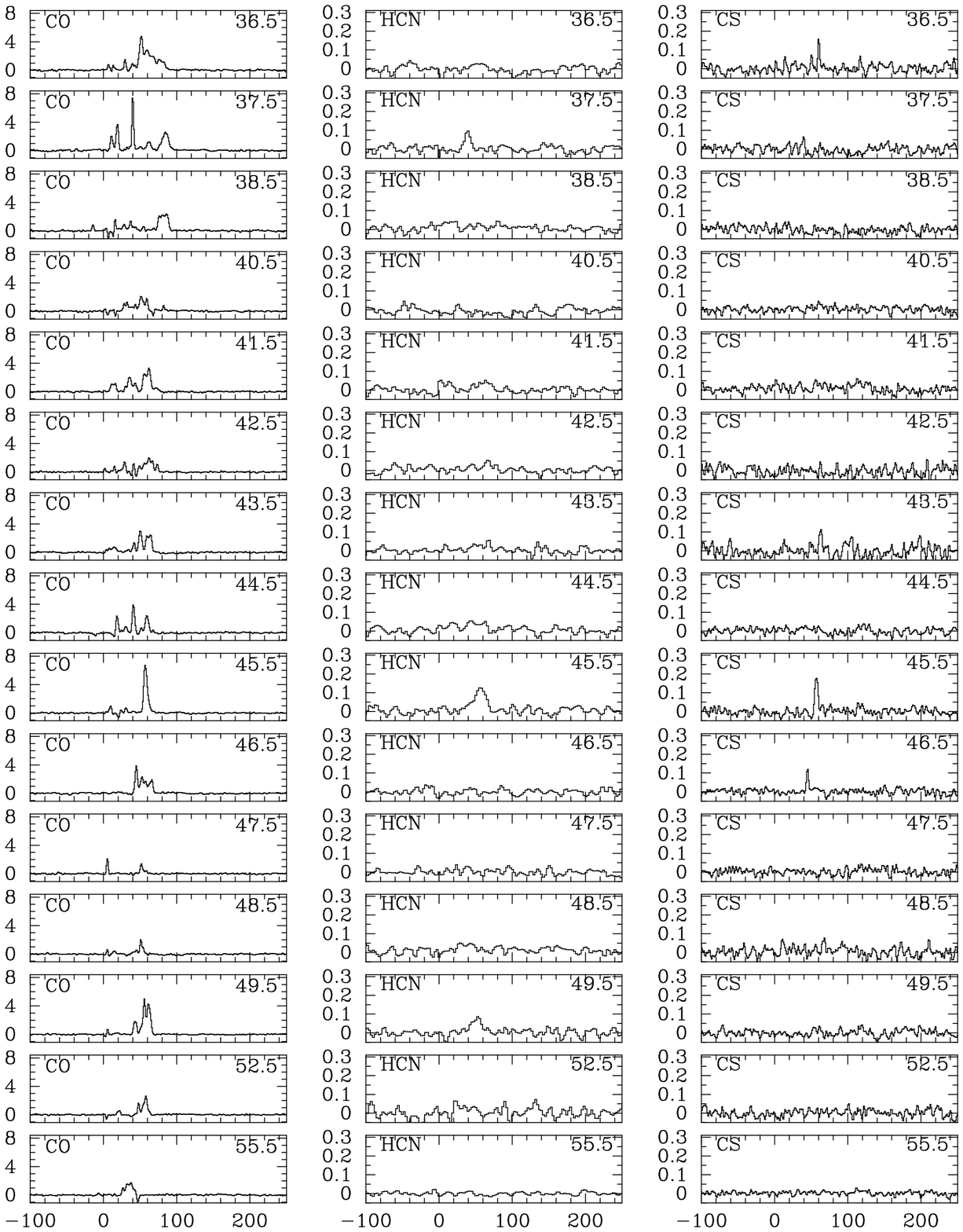}{460pt}{0}{80}{80}{-248}{-72}
\caption{continued. }
\end{figure}

\clearpage


\begin{thebibliography}{}

\bibitem{}
Bally, J., Stark, A.A., Wilson, R.W., \& Henkel, C. 1987, \apjs, 65, 13
\bibitem{} Blitz, L. 1987, in Physical Processes in Interstellar Clouds,
ed. G.E. Morfill \& M. Scholer (Dordrecht:Reidel), 35
\bibitem{} 
Blitz, L. 1993, in Protostars and Planets III, ed. E.H. Levy 
\& J.I. Lunine (Tucson:Univ. of Arizona Press), 12
\bibitem{}
Dame, T.M. 1993, in Back to the Galaxy, eds. S.S. Holt \& F. Verter 
(New York:AIP), AIP Conference Proceedings 278, 267
\bibitem{}
Dame, T.M., Ungerechts, H., Cohen, R.S., de Geus, E.J., Grenier, I.A., May, J.,
Murphy, D.C., Nyman, L.-\AA., \& Thaddeus, P. 1987, \apj, 322, 706
\bibitem{}
Digel, S., de Geus, E., \& Thaddeus, P. 1994, \apj, 422, 92
\bibitem{}
Gould, A. 1996, in Unsolved Problems of the Milky Way, eds. L. Blitz
\& P. Teuben (Dordrecht:Kluwer), 651
\bibitem{}
G\"{u}sten, R. 1989, in The Center of the Galaxy, ed. Morris, M. 
(Dordrecht:Kluwer), 89
\bibitem{}
Handa, T., Hasegawa, T., Hayashi, M., Sakamoto, S., Oka, T., \& Dame, T.M.
1993, in Back to the Galaxy, eds. S.S. Holt \& F. Verter
(New York:AIP), AIP Conference Proceedings 278, 315
\bibitem{}
Helfer, T.T. \& Blitz, L. 1993, \apj, 419, 86
\bibitem{}
Jackson, J.M., Heyer, M.H., Paglione, T.A.D., \& Bolatto, A.D. 1996, \apjlett, 
456, 91
\bibitem{}
Lada, E.A., Bally, J., \& Stark, A.A. 1991, \apj, 368, 432
\bibitem{}
Liszt, H.S. 1993, \apj, 411, 720
\bibitem{}
Liszt, H.S. 1995, \apj, 442, 163
\bibitem{}
Liszt, H.S., Burton, W.B., \& Xiang, D.L. 1984, \aap, 140, 303
\bibitem{}
Magnani, L., Blitz, L., \& Mundy, L.G. 1985, \apj, 295, 402
\bibitem{} 
Nguyen-Q-Rieu, Jackson, J. M., Henkel, C., Truong-Bach, \&
Mauersberger, R. 1992, \apj, 399, 521
\bibitem{}
Sakamoto, S., Hasegawa, T., Hayashi, M., Handa, T., \& Oka, T. 1995,
\apjsupp, 100, 125
\bibitem{}
--------------. 1996, submitted to ApJ
\bibitem{}
Sanders, D.B., Scoville, N.Z., Tilanus, R.P.J., Wang, Z., \& Zhou, S.
1993, in Back to the Galaxy, eds. S.S. Holt \& F. Verter
(New York:AIP), AIP Conference Proceedings 278, 311
\bibitem{}
Sanders, D.B., Solomon, P.M., \& Scoville, N.Z. 1984, \apj, 276, 182
\bibitem{}
Sault, R.J., Teuben, P.J., \& Wright, M.C.H. 1995, in Astronomical Data 
Analysis Software and Systems IV, eds. R.A. Shaw, H.E. Payne, \& J.J.E. Hayes, 
A.S.P.  Conference Series 77, 433 
\bibitem{} 
Spergel, D.N. \& Blitz, L. 1992, Nature, 357, 665
\bibitem{}
Spergel, D.N., Malhotra, S., \& Blitz, L. 1996, in Spiral Galaxies in 
the Near Infrared, eds. D. Minnitti \& H.-W. Rix, in press
\bibitem{}
Waller, W.H. \& Tacconi-Garman, L.E. 1992, \apjs, 80 305
\bibitem{}
Williams, J.P. 1995, Ph.D. thesis, University of California

\end{thebibliography}
\end{document}